\documentclass[showpacs,preprintnumbers,amsmath,amssymb]{revtex4}
\usepackage{amsmath,amssymb,graphics,epsfig,subfigure}
\usepackage{color}

\begin{document}
\newcommand {\nn} {\nonumber}
\renewcommand{\baselinestretch}{1.3}

\title{Triple points and phase diagrams in the extended phase space of charged Gauss-Bonnet black holes in AdS space}

\author{Shao-Wen Wei \footnote{weishw@lzu.edu.cn},
        Yu-Xiao Liu \footnote{liuyx@lzu.edu.cn}}

\affiliation{Institute of Theoretical Physics, Lanzhou University, Lanzhou 730000, People's Republic of China}

\begin{abstract}
We study the triple points and phase diagrams in the extended phase space of the charged Gauss-Bonnet black holes in $d$-dimensional anti-de Sitter space, where the cosmological constant appears as a dynamical pressure of the system and its conjugate quantity is the thermodynamic volume of the black holes. Employing the equation of state $T=T(v, P)$, we demonstrate that the information of the phase transition and behavior of the Gibbs free energy are potential encoded in the $T-v$ ($T-r_{h}$) line with fixed pressure $P$. We get the phase diagrams for the charged Gauss-Bonnet black holes with different values of the charge $Q$ and dimension $d$. The result shows that the small/large black hole phase transitions appear for any $d$, which is reminiscent of the liquid/gas transition of a van der Waals type. Moreover, the interesting thermodynamic phenomena, i.e., the triple points and the small/intermediate/large black hole phase transitions are observed for $d=6$ and $Q\in(0.1705, 0.1946)$.
\end{abstract}



\maketitle

\section{Introduction}
\label{secIntroduction}

Black holes are now widely believed to be thermodynamic objects assigned standard thermodynamic variables such as the temperature and entropy \cite{Bardeen,Bekenstein}. A lot of attention has been devoted to studying thermodynamic properties for different black holes. Motivated by the AdS/CFT correspondence \cite{Maldacena,Gubser,Witten} that thermodynamics of a black hole in AdS space can be identified with that of dual strongly coupled conformal field theory on the boundary of the AdS space, the thermodynamic properties of black holes in anti-de Sitter (AdS) space are researched recently. In AdS space, the so-called Hawking-Page phase transition can occur between stable large black holes and thermal gas in AdS space \cite{Hawking}, which can well explain the confinement/deconfinement phase transition of gauge field \cite{Hawking,Witten2}.

A decade ago, thermodynamics of the charged black holes in AdS space was studied. It was shown in Refs. \cite{Chamblin,Chamblin2,Lemos} that, in a canonical ensemble the black hole system has a small/larger black hole (SBH/LBH) phase transition. With the increasing of the charge, such phase transition terminates at a critical point. The $Q-\Phi$ (charge-chemical potential) diagram of the charged black holes is quite similar to the $P-v$ (pressure-volume) diagram of the van der Waals fluid \cite{Shen,Kubiznak,Wei,Niu,Tsai,Banerjee,Banerjee3,Lala,Spallucci,Smailagic}. Therefore the black hole system is shown to be analogous to the van der Waals fluid. Although the analogy between them is widely studied, it is somewhat problematic because that $Q$ and $\Phi$ are extensive and intensive quantities, while $P$ and $v$ are intensive and extensive quantities. On the other hand, such phase transition of van der Waals type can be also observed for asymptotically flat or de Sitter charged black holes or black branes placed in a finite cavity \cite{Vaidya,Lu,Roy}.

Very recently, by interpreting the cosmological constant as a thermodynamic pressure and its conjugate quantity as a thermodynamic volume of the black hole \cite{Dolan,Cvetic,Kastor,Traschen,Castro,Menoufi}, the analogy between the charged black hole in AdS space and van der Waals fluid was further enhanced \cite{Kubiznak}. Both systems were found to have the same critical exponents near the critical point and extremely similar phase diagrams. The problematic between $Q-\Phi$ and $P-v$ was modified. Thus the analogy between the charged AdS black hole and the van der Waals system becomes more complete. This analogy has been generalized to different charged black holes and rotating black holes in AdS space in the extended phase space \cite{Gunasekaran,Hendi,Chen,ZhaoZhao,Altamirano,Cai,AltamiranoKubiznak,XuXu,Mo,zou,MoLiu,Altamirano3,Roychowdhury}.

Among these studied, it was shown that besides the SBH/LBH first-order phase transition reminiscent of the liquid/gas transition of the van der Waals fluid, there also appear some new interesting phenomena analogous to the ``every day thermodynamics" of simple substances, such as reentrant phase transitions of multicomponent liquids, multiple first-order solid/liquid/gas phase transitions, and liquid/gas phase transitions of the van der Waals type. It was shown that in all $d\geq 6$ dimensions the single spinning vacuum Kerr-AdS black holes demonstrate the peculiar behavior of LBH/SBH/LBH phase transitions reminiscent of reentrant phase transitions \cite{Altamirano}. Such reentrant phase transition was also found in the four-dimensional Born-Infeld-AdS black hole spacetimes, deep in the nonlinear regime of the Born-Infeld theory. However, for the higher dimensional Born-Infeld AdS black holes, such phase transition was not observed \cite{zou}. More intriguingly, the multiply rotating Kerr-AdS black hole system displays a small/intermediate/large black hole (SBH/IBH/LBH) phase transition with one tricritical and two critical points in some range of the parameters \cite{AltamiranoKubiznak}. It was argued in Refs. \cite{Altamirano,AltamiranoKubiznak} that the new phase structure in the thermodynamics of rotating black holes resembles to binary fluids seen in superfluidity and superconductivity. More recently, the authors in Ref. \cite{Altamirano3} also pointed out that the reentrant phase transitions are observed for the asymptotically flat doubly-spinning Myers-Perry black holes of vacuum Einstein gravity. Hence, neither exotic matter nor a cosmological constant is required for this phenomenon to occur in black hole spacetimes.

The aim of this paper is to search for the possible triple points and SBH/IBH/LBH black hole transitions in the charged Gauss-Bonnet (GB) black holes in AdS space. In fact, in our previous work \cite{Wei}, we studied the $Q-\Phi$ criticality for the charged GB black holes. We ignored a subtle thing that worthwhile to note here: we found that the black hole system undergoes two phase transitions of the van der Waals type at one isotherm in some ranges of the parameter (see Table I in Ref. \cite{Wei} for details). This in other words implies a reentrant phase transition or a triple point. It is the partial aim of this paper to examine such phase transition. Considering the problematic of the analogy between the analogy of $Q-\Phi$ and $P-v$, we will do it in the extended phase space, where the cosmological constant appears as a dynamical pressure of the system and its conjugate quantity is the thermodynamic volume of the black holes. The result indicates that there indeed exists the SBH/IBH/LBH black hole transition in the case of $d=6$ and $Q\in(0.1705, 0.1946)$. And the triple point is also found, at which the SBH, IBH, and LBH can coexist. For the other values of the parameters, the phase diagrams are also obtained, which reveals that the SBH/LBH phase transition is ubiquitous.

The outline of our paper is as follows. In Sec. \ref{vanderWaalsfluid}, we review the van der Waals fluid. Through rewriting the equation of state in the form of $T=T(v, P)$, we show that the critical behavior and the phase diagram can be equivalently obtained. In Sec. \ref{CGB}, we give some thermodynamic quantities of the charged GB black holes in AdS space, and we clearly show that the information of the phase transition is encoded in the $T-r_{h}$ line with fixed $P$. The divergence and sign of the heat capacity are also encoded in it. In Secs. \ref{Phase1}, \ref{Phase2}, and \ref{Phase3}, we respectively discuss the phase diagrams of $d=5$, $d=6$, and $d\geq 7$ in detail. Section \ref{Conclusion} is devoted to the conclusions and discussions.

\section{Review of van der Waals fluid}
\label{vanderWaalsfluid}

Compared to the ideal gas, the van der Waals fluid approaches the real fluid for considering the nonzero size of molecules and the attraction between them. The equation of state and the Gibbs free energy read
\begin{eqnarray}
 T&=&\bigg(P+\frac{a}{v^{2}}\bigg)(v-b),\label{pt0}\\
 G&=&-T\bigg(1+\ln (v-b)T^{3/2}/\Phi_{0}\bigg)-\frac{a}{v}+Pv,
\end{eqnarray}
where we have set the Boltzmann constant $k=1$. $v=V/N$, $P$, $T$, and $\Phi_{0}$ are the specific volume of the fluid, pressure, temperature, and a (dimensionful) constant characterizing the gas, respectively. The parameter $a$ measures the attraction between the molecules, and $b$ describes the nonzero size of the molecules.

\begin{figure}
\includegraphics[width=8cm]{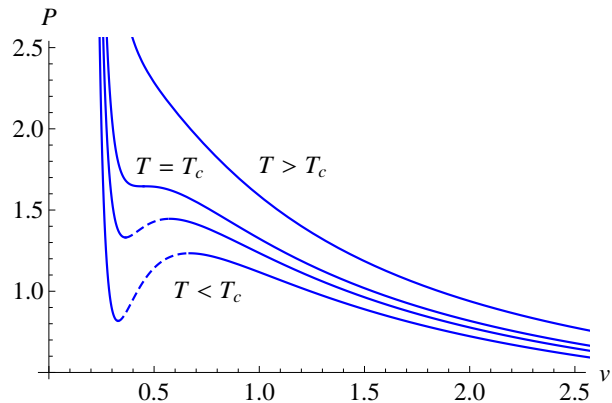}
\caption{$P-v$ diagram of van der Waals fluid for fixed $T$. The parameters $a=1$, $b=0.15$. The temperature $T=2.2$, $T_{c}$, 1.9, and 1.8 from top to bottom. Here the critical temperature $T_{c}=1.9753$.}\label{Pvanpv}
\end{figure}

\begin{figure}
\center{\subfigure[]{\label{PCriticalPb}
\includegraphics[width=8cm,height=6cm]{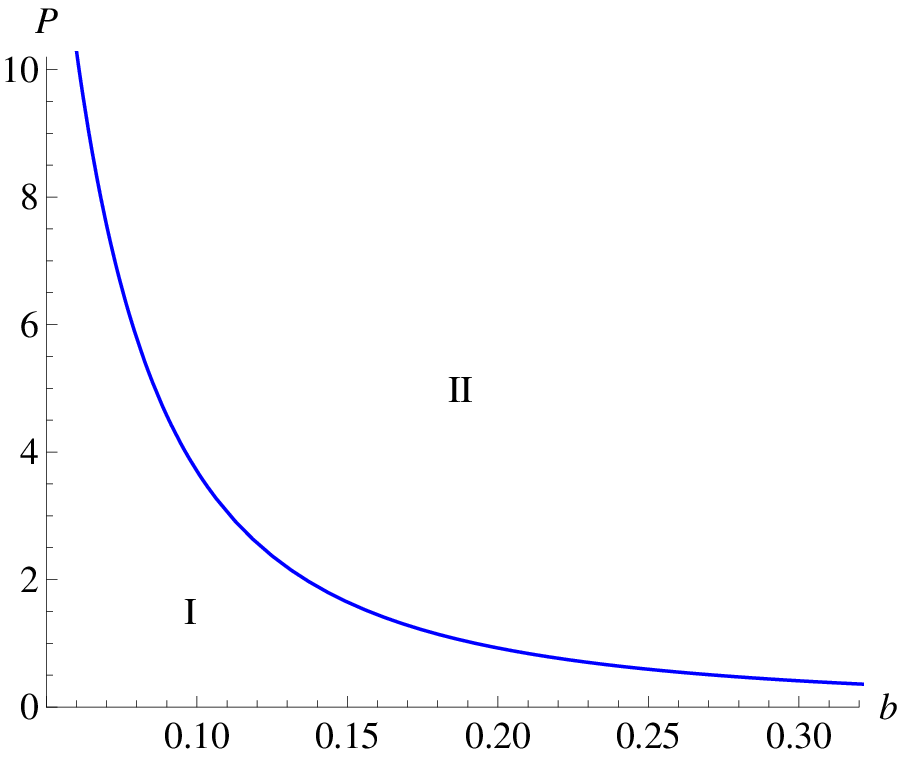}}
\subfigure[]{\label{Ppvt}
\includegraphics[width=8cm,height=6cm]{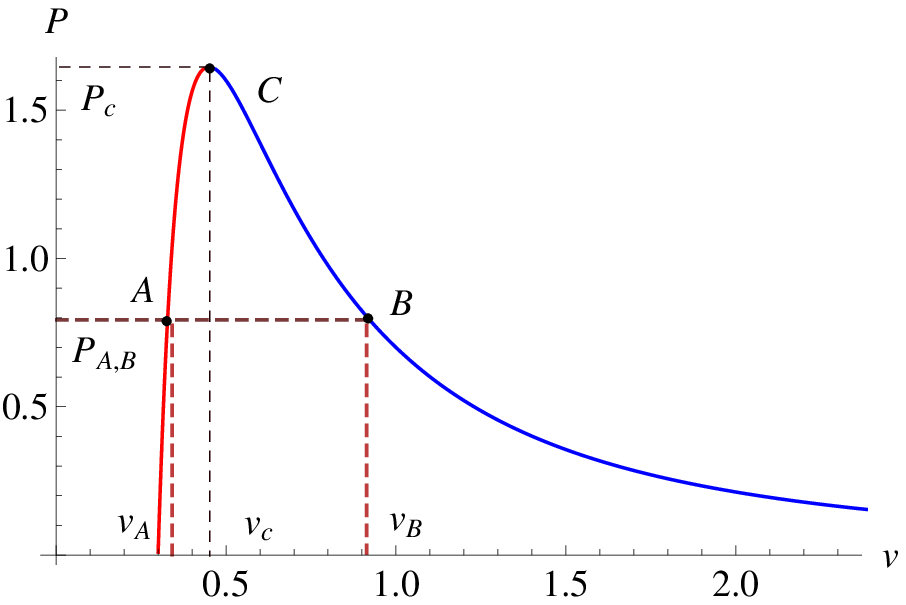}}}
\caption{The parameter $a=1$. (a) The critical point in $P-b$ plane. The equation $(\partial_{v}T)_{P}=0$ has two roots in the range I, one root at the line, and no root in the range II. (b) Roots of equation $(\partial_{v}T)_{P}=0$ with $b=0.15$.}
\end{figure}

\begin{figure}
\center{\subfigure[]{\label{Ptv}
\includegraphics[width=8cm,height=6cm]{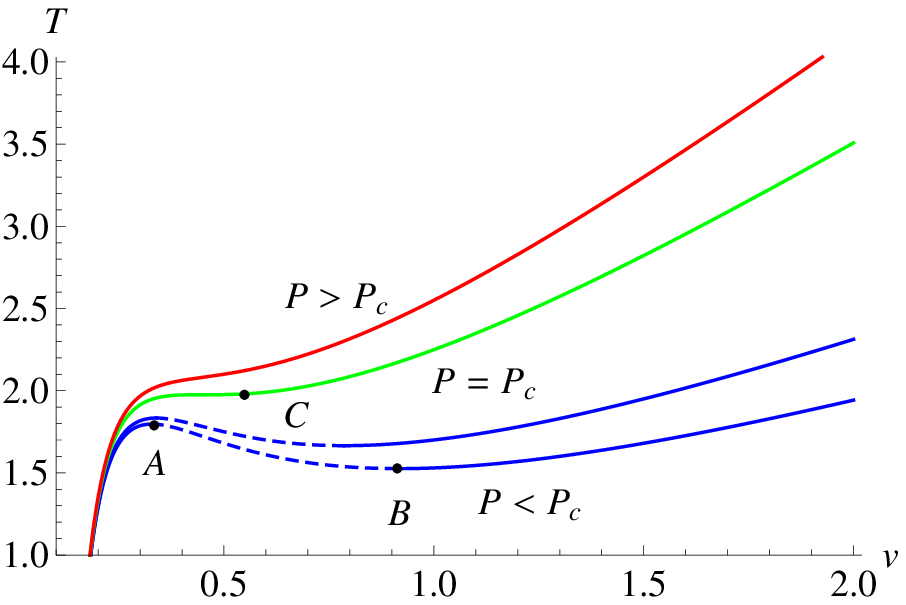}}
\subfigure[]{\label{Pgt}
\includegraphics[width=8cm,height=6cm]{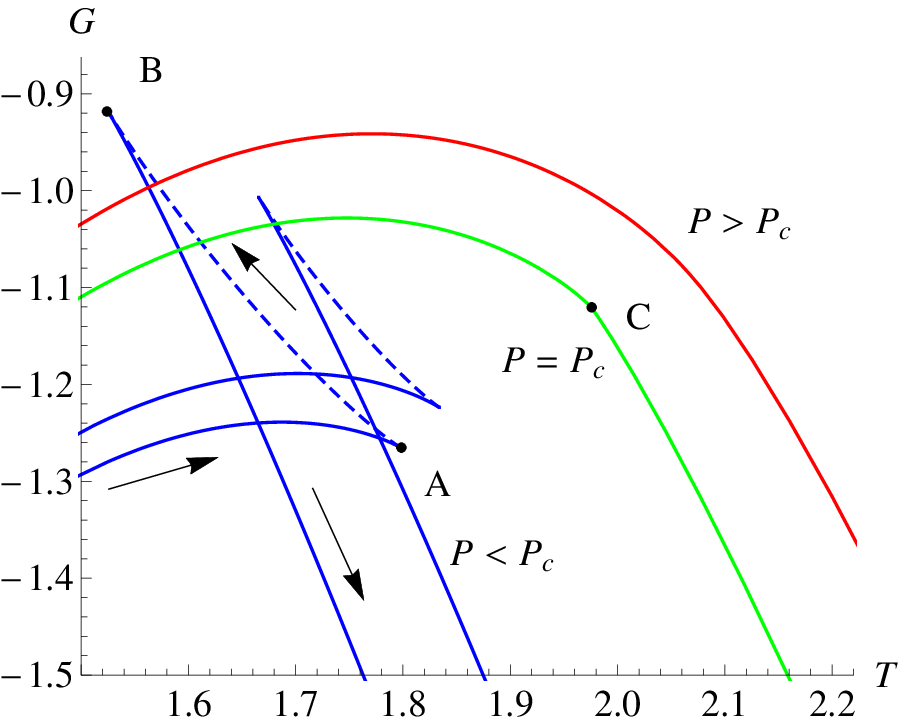}}}
\caption{(a) The temperature $T$ as a function of the volume $v$. (b) The Gibbs free energy as a function of the temperature $T$ with $\Phi_{0}=1$. The parameters $a=1$, $b=0.15$. And the pressure $P=2$, $P_{c}$, 1, and 0.8 from top to bottom, with $P_{c}=1.6461$. Black arrows indicate increasing $v$.}\label{Pv2}
\end{figure}

The state function (\ref{pt0}) (reexpressed in the form $P=P(v, T)$) is often used to describe the basic qualitative features of the liquid/gas phase transition. The qualitative behavior of isotherms in the $P-v$ diagram for the van der Waals fluid can be found in Fig. \ref{Pvanpv}. From it, we clear see that there exists a critical case at $T=T_{c}$. For $T>T_{c}$, it describes the ``ideal gas''. And for $T<T_{c}$, there is a classical behavior of isotherms for the van der Waals fluid. There are three branches in such case, two have negative slope described by the solid lines. While the third one has positive slope described in the dashed line, which is unphysical and should be replaced by the horizontal line determined by the Maxwell's equal area law. Such behavior generally implies a first-order liquid/gas phase transition. The critical point is an inflection point, which is determined by
\begin{eqnarray}
 \bigg(\frac{\partial P}{\partial v}\bigg)_{T}=0,\quad
 \bigg(\frac{\partial^{2} P}{\partial v^{2}}\bigg)_{T}=0.
\end{eqnarray}
The index means the temperature $T$ is fixed. Solving this equation, we obtain the critical point
\begin{eqnarray}
 T_{c}=\frac{8a}{27b},\quad v_{c}=3b,\quad
 P_{c}=\frac{a}{27b^{2}}.\label{vdwcp}
\end{eqnarray}
And at the critical point
\begin{eqnarray}
 P_{c}v_{c}/T_{c}=3/8,
\end{eqnarray}
which is a universal number for all fluids. On the other hand, the equation of state (\ref{pt0}) can be regarded as $T=T(v, P)$. And since $(\partial_{P}T)_{v}=(v-b)>0$, the critical point can be equivalently determined by
\begin{eqnarray}
 \bigg(\frac{\partial T}{\partial v}\bigg)_{P}=0,\quad
 \bigg(\frac{\partial^{2} T}{\partial v^{2}}\bigg)_{P}=0.\label{vdwcp2}
\end{eqnarray}
Solving it, the critical point (\ref{vdwcp}) will be obtained. The critical point satisfying (\ref{vdwcp2}) is depicted in Fig. \ref{PCriticalPb} with $a=1$. After a simple calculation, the equation $(\partial_{v}T)_{P}=0$ has two roots in the range I, one root at the line, and no root in the range II. For an example, we plot the roots of $(\partial_{v}T)_{P}=0$ in Fig. \ref{Ppvt}, for $a=1$ and $b=0.15$. From it, we see that there exists a maximal value of $P$ corresponding to the critical pressure $P_{c}$, which can be found in Fig. \ref{PCriticalPb} with $b=0.15$. For small $P<P_{c}$, there are two roots. And the two roots meet each other at $P=P_{c}$. With $P$ further increasing, no root can be found. The corresponding behavior of $T$ as a function of $v$ with fixed $P$ is given in Fig. \ref{Ptv}. For $P>P_{c}$, $T$ monotonically increases with $v$, which denotes the ``ideal gas" branch. However, when $P<P_{c}$, the two roots showed in Fig. \ref{Ppvt}, i.e., the points $A$ and $B$, divide one line into three branches. Two of the branches in the solid lines have positive slope, while the other one in the dashed line has negative slope. The Gibbs free energy is depicted in Fig. \ref{Pgt} with the same parameters. A detailed study shows that the three branches in the $P-v$ line with constant $P<P_{c}$ are consistent with the three branches in the $G-T$ line with the same constant $P$. The discontinuous points in $G-T$ line are the roots of $(\partial_{v}T)_{P}=0$ (see the points $A$ and $B$ shown in Fig. \ref{Ppvt} and Fig. \ref{Pv2}). It is clear that the characteristic swallow tail behavior of the Gibbs free energy appears for $P<P_{c}$ implies a first-order phase transition. The coexistence line of liquid and gas phases of the van der Waals fluid can be determined by the intersection of the two branches of the same $G-T$ line shown in solid line, and we do not show it here.

Now to summarize the result. The appearance of the swallow tail behavior implies a first-order phase transition. And the behavior needs at least three branches on one $G-T$ line, which is bounded with two discontinuous points determined by $(\partial_{v}T)_{P}=0$. So if there is no or only one discontinuous point, then there is no swallow tail behavior, and therefore there is no the liquid/gas phase transition of van der Waals type. In general, such phase transition is studied with the state function $P=P(v, T)$. However, we show in the above that with the state function expressed in $T=T(v, P)$, the phase transition can also be equivalently obtained. In the following, we will show that for a black hole system, $T=T(v, P)$ is a natural and convenient choice. Under such choice, we can get some information of the phase transition through examining the behavior of the temperature $T$ vs. pressure $v$. Moreover, we will show in the next section that the discontinuous point in the $G-T$ line has a local extremum in the $T-v$ ($T-r_{h}$) line. And the points is related to a divergence of the heat capacity. Adopting such choice of the equation of state, we will investigate the phase diagram of the charged GB black holes in AdS space.

\section{Thermodynamics of charged Gauss-Bonnet black holes in AdS space}
\label{CGB}

The action of the Einstein-Gauss-Bonnet gravity in $d$-dimensional spacetime with a negative cosmological constant reads as
\begin{eqnarray}
 S=\int d^{d}x\sqrt{-g}
 \Big[\frac{1}{16\pi G_{d}}(\mathcal{R}
    -2\Lambda
    +\alpha_{\text{GB}}\mathcal{L}_{\text{GB}})
  -\mathcal{L}_{\text{matter}}\Big], \label{action}
\end{eqnarray}
where $\alpha_{\text{GB}}$ is the GB coupling constant with dimension $(\text{length})^{2}$ and it is regarded as the inverse string tension with positive value. The GB Lagrangian $\mathcal{L}_{\text{GB}}$ and the electromagnetic Lagrangian $\mathcal{L}_{matter}$ are
\begin{eqnarray}
&&\mathcal{L}_{\text{GB}}=\mathcal{R}_{\mu\nu\gamma\delta}\mathcal{R}^{\mu\nu\gamma\delta}
                    -4\mathcal{R}_{\mu\nu}\mathcal{R}^{\mu\nu}+\mathcal{R}^{2},\\
&&\mathcal{L}_{\text{matter}}=4\pi \mathcal{F}_{\mu\nu}\mathcal{F}^{\mu\nu}.
\end{eqnarray}
The Maxwell field strength is defined as
$\mathcal{F}_{\mu\nu}=\partial_{\mu}\mathcal{A}_{\nu}-\partial_{\nu}\mathcal{A}_{\mu}$
with $\mathcal{A}_{\mu}$ the vector potential. The solution of the $d$-dimensional static charged GB-AdS black hole for the action (\ref{action}) is
\begin{eqnarray}
 ds^{2}=-f(r)dt^{2}+f^{-1}(r)dr^{2}+r^{2}(d\theta^{2}+\sin^{2}\theta d\phi^{2}+\cos^{2}\theta d\Omega^{2}_{d-4}),\label{metric}
\end{eqnarray}
with the metric function given by \cite{Boulware,Cai2,Wiltshire,Cvetic2}
\begin{eqnarray}
 f(r)=1+\frac{r^{2}}{2\alpha}
   \left[1-\sqrt{1 + \frac{2\alpha}{d-2} \left(\frac{32\pi M}{\Sigma_{d-2} r^{d-1}}-\frac{4 Q^{2}}{(d-3)r^{2d-4}}+\frac{4\Lambda}{d-1}\right)}\right],
\end{eqnarray}
where $\alpha=(d-3)(d-4)\alpha_{\text{GB}}$. $\Sigma_{d-2}$ is the area of a unite $(d-2)$-dimensional sphere, and we set $\Sigma_{d-2}=1$ for simplicity. Recent development on the thermodynamics of black holes in extended phase space shows that the cosmological constant can be interpreted as the thermodynamic pressure and treated as a thermodynamic variable in its own right,
\begin{eqnarray}
 P=-\frac{1}{8\pi}\Lambda.
\end{eqnarray}
It is shown that the differential form holds \cite{Cai}
\begin{eqnarray}
 dH=TdS+\Phi dQ+vdP+\mathcal{A}d\alpha,
\end{eqnarray}
where $H\equiv M$ is the enthalpy of the gravitational system \cite{KastorRay}. The thermodynamic volume $v$ is the thermodynamic quantity conjugating to the pressure $P$. And $\mathcal{A}$ is the conjugate quantity conjugating to the GB coefficient $\alpha$. We list the thermodynamic quantities
\begin{eqnarray}
 H&=&\frac{(d-2)r_{h}^{d-3}}{16\pi}\bigg(1+\frac{\alpha}{r_{h}^{2}}+
     \frac{16\pi Pr_{h}^{2}}{(d-1)(d-2)}\bigg)+\frac{Q^{2}}{8\pi(d-3)r_{h}^{d-3}},\\
 T&=&\frac{16\pi Pr_{h}^{4}/(d-2)+(d-3)r_{h}^{2}+(d-5)\alpha-2Q^{2}r_{h}^{8-2d}/(d-2)}
    {4\pi r_{h}(r_{h}^{2}+2\alpha)},\\
 S&=&\frac{r_{h}^{d-2}}{4}\left(1+\frac{2(d-2)\alpha}{(d-4)r_{h}^{2}}\right),
     \quad v=\frac{r_{h}^{d-1}}{d-1},\quad \Phi=\frac{Q}{4\pi(d-3)r_{h}^{d-3}},\\
 \mathcal{A}&=&\frac{(d-2)}{16\pi}r_{h}^{d-5}-\frac{(d-2)T}{2(d-4)}r_{h}^{d-4},\\
 16\pi G&=&(d-2) r_{h}^{d-3}
    \bigg(1+\frac{1}{r_{h}^2}+\frac{16 \pi  P r_{h}^2}{d^2-3d+2}\bigg)
    +\frac{2 Q^2 r_{h}^{3-d}}{d-3}\nonumber\\
   &&-\frac{r_{h}^{d-3}}{2+r_{h}^2}
   \bigg(1+\frac{2(d-2)}{(d-4)r_{h}^2}\bigg)
   \bigg(d-5+(d-3) r_{h}^2+\frac{16 \pi  P r_{h}^4-2 Q^2 r_{h}^{8-2
   d}}{d-2}\bigg).\label{quantities}
\end{eqnarray}
$r_{h}$ is the horizon radius of the black hole and it is determined by the largest real root of the equation $f(r_{h})=0$. The range of $T<0$ represents the non-black hole case. So, we will only focus on the range of $T>0$ and $S>0$ in this paper.

The heat capacity $C_{P}$ at fixed $P$ measuring the local thermodynamical stability is defined as
\begin{eqnarray}
 C_{P}=T\bigg(\frac{\partial S}{\partial T}\bigg)_{P}.
\end{eqnarray}
Using the thermodynamic quantities, $C_{P}$ can be reexpressed as
\begin{eqnarray}
 C_{P}=T\bigg(\frac{\partial_{r_{h}} S}{\partial_{r_{h}} T}\bigg)_{P}
       \propto (\partial_{r_{h}} T)^{-1}_{P}.\label{heat}
\end{eqnarray}
The second step is guaranteed with $(\partial_{r_{h}} S)_{P}>0$ for $d\geq5$ and positive temperature for a black hole system. Thus it is qualitatively clear that the heat capacity $C_{P}$ diverges at $(\partial_{r_{h}} T)_{P}=0$. And positive (negative) $(\partial_{r_{h}} T)^{-1}_{P}$ relates with positive (negative) $C_{P}$. Therefore, in the $T-r_{h}$ plane, the black hole branch of (negative) positive slope relate to the thermodynamically (unstable) stable phase. Since $(\partial_{P}T)_{v}>0$ for positive $\alpha$, the critical point can be determined by Eq. (\ref{vdwcp2}). On the other hand we have $v\propto r_{h}^{d-1}$, so the critical condition (\ref{vdwcp2}) for the black hole can be expressed as
\begin{eqnarray}
 \bigg(\frac{\partial T}{\partial r_{h}}\bigg)_{P}=0,\quad
 \bigg(\frac{\partial^{2} T}{\partial r_{h}^{2}}\bigg)_{P}=0.\label{vdwcp3}
\end{eqnarray}
In the following discussion, without loss of generality we adopt $\alpha=1$. Several remarks on the $T-r_{h}$ line and phase transitions are in order:

(i) The extremal points determined by $(\partial_{r_{h}}T)_{P}=0$ divide one $T-r_{h}$ line into several black hole branches. The extremal points correspond to these discontinuous points in the $G-T$ line.

(ii) At the extremal points, the heat capacity diverges.

(iii) The black hole branch with (negative) positive slope in the $T-r_{h}$ plane is thermodynamically (unstable) stable.

(iv) The critical point occurs at the multiple root of $(\partial_{r_{h}}T)_{P}=0$.

(v) The number of the extremal points could tell us the number of the characteristic swallow tail behavior of the Gibbs free energy, and the possible existence of the phase transition of ``everyday thermodynamics". Swallow tail behavior and phase transition will not occur if there is no or one extremal point, while two extremal points generally produce one swallow tail behavior and the liquid/gas phase transitions of the van der Waals type. More extremal points will potentially give the reentrant phase transitions.

\section{The phase diagram when $d=5$}
\label{Phase1}

\begin{figure}
\center{\subfigure[]{\label{Pd5PQ}
\includegraphics[width=8cm,height=6cm]{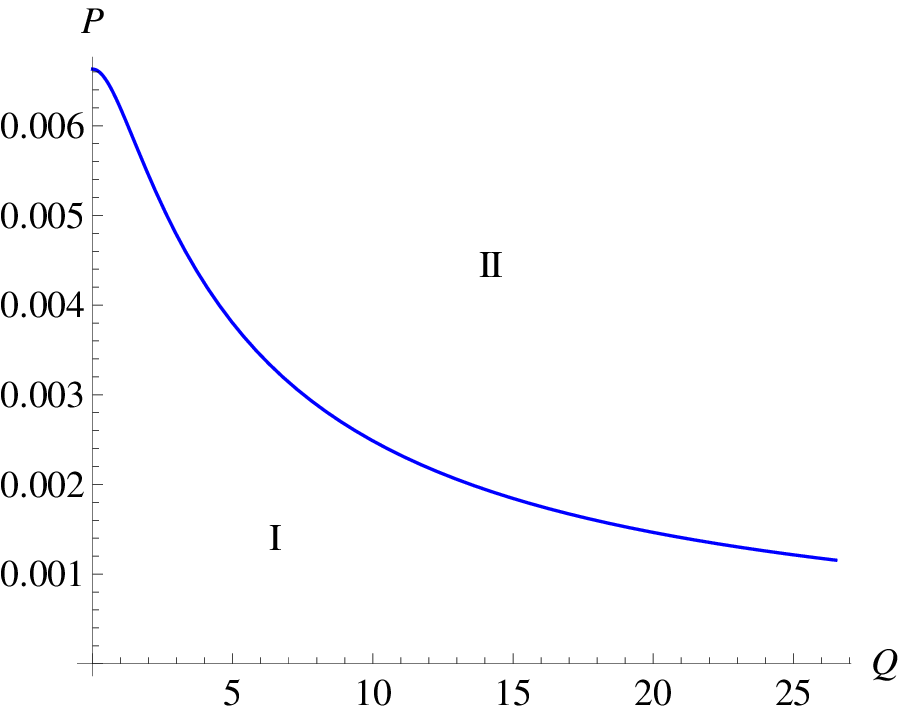}}
\subfigure[]{\label{Pd5CP}
\includegraphics[width=8cm,height=6cm]{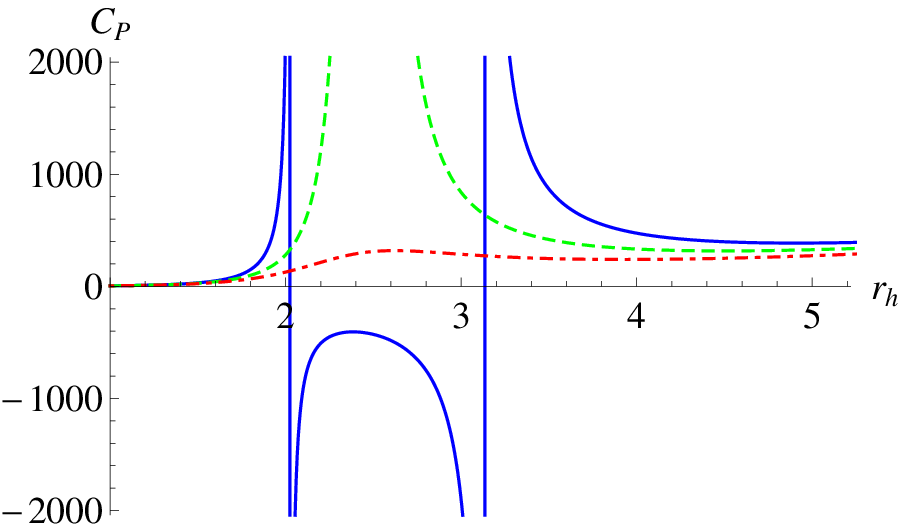}}}
\caption{In five-dimensional spacetime. (a) The critical point displayed in $P-Q$ plane. (b) The heat capacity $C_{P}$ as a function of $r_{h}$ with fixed $Q$=0.18. The blue solid line, green dashed line, and red dot dashed lines are for $P$=0.0060, 0.0066, and 0.0075, respectively.}
\end{figure}

\begin{figure}
\center{\subfigure[]{\label{Pd5Trh}
\includegraphics[width=8cm,height=6cm]{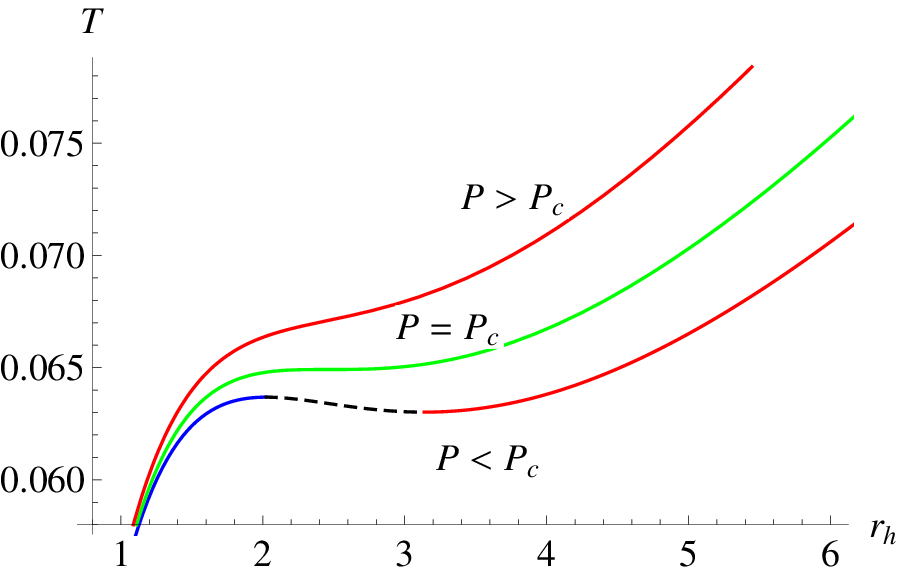}}
\subfigure[]{\label{Pd5GT}
\includegraphics[width=8cm,height=6cm]{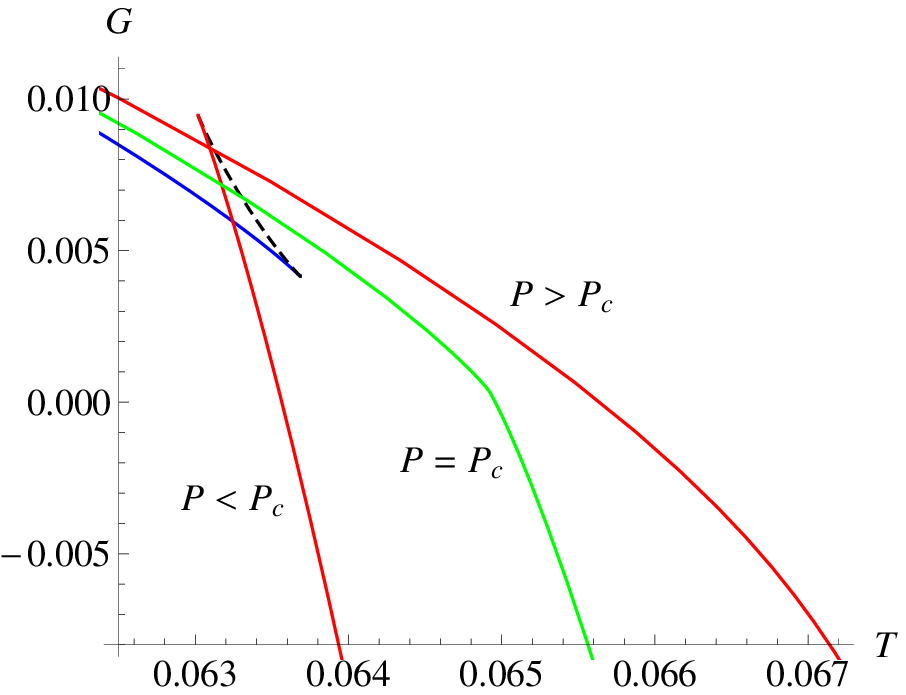}}}
\caption{In five-dimensional spacetime and $Q=0.18$, $P=$0.0060, 0.0066, and 0.0075 from bottom to top. The branches described by the dashed line have negative heat capacity $C_{P}$ indicating locally thermodynamically unstable. (a) Behavior of $T$ as a function of $r_{h}$. For $P<P_{c}$, there are one unstable and two stable branches. (b) The Gibbs free energy as a function of $T$. The characteristic swallow tail behavior appears for $P<P_{c}$. The horizon radius $r_{h}$ increases from left to right along the $G-T$ line.}\label{PPrhgt}
\end{figure}

For a five-dimensional static charged GB-AdS black hole ($d=5$), the temperature, Gibbs free energy, and heat capacity are
\begin{eqnarray}
 T&=&\frac{8 \pi Pr_{h}^6+3r_{h}^4-Q^2}{6\pi r_{h}^5+12\pi r_{h}^3},\\
 G&=&\frac{-4\pi Pr_{h}^8+(3-72\pi P)
   r_{h}^6+Q^2\left(5r_{h}^2+18\right)-9r_{h}^4+18r_{h}^2}{48\pi r_{h}^2
   \left(r_{h}^2+2\right)},\\
 C_{P}&=&\frac{3 r_{h}\left(r_{h}^2+2\right)^2
                \left(8\pi Pr_{h}^6-Q^2+3 r_{h}^4\right)}
              {4\left[r_{h}^4 \left(8\pi Pr_{h}^4
                +(48\pi P-3)r_{h}^2+6\right)+Q^2 \left(5r_{h}^2+6\right)\right]}.
\end{eqnarray}
The critical point can be found with the condition (\ref{vdwcp3}). In the small charge limit, we have
\begin{eqnarray}
 \frac{P_{c}r_{hc}}{T_{c}}=\frac{1}{4}+\frac{1}{72}Q^{2}+\mathcal{O}(Q^{4}).
\end{eqnarray}
This result is different from that of the van der Waals fluid, for which the right-hand side of the corresponding equation is $8/3$. When the charge $Q$=0, our result recovers that shown in Ref. \cite{Cai}. The behavior of the critical point is displayed in the $P-Q$ plane in Fig. \ref{Pd5PQ}. It behaves very similar to that of van der Waals fluid shown in Fig. \ref{PCriticalPb}. In the region I, $(\partial_{r_{h}}T)_{P}=0$ has two roots, one is the local maximum and another is the local minimum. On the line, the two roots meet each other. And in the region II, there exists no root. However, different from the van der Waals fluid, the critical pressure $P_{c}$ here has a maximum $1/(48\pi)$ at $Q=0$. The heat capacity is plotted in Fig. \ref{Pd5CP} for $Q=0.18$. When $P=0.0060<P_{c}$, $(\partial_{r_{h}}T)_{P}=0$ (described by the blue solid line) has two roots, at which $C_{P}$ diverges. It is also clear that $C_{P}$ changes its sign at the roots. When $P=P_{c}\approx0.0066$, the two roots coincide with each other, resulting only one divergent behavior of $C_{P}$ described by the green dashed line, and the range of negative $C_{P}$ is squeezed out. Further increase the pressure to $P=0.0075$, $C_{P}$ will be positive and continuous for any $r_{h}$. For the three constant values of the pressure $P$, i.e., $P$=0.0060, 0.0066, and 0.0075, the behavior of the temperature $T$ is depicted in Fig. \ref{Pd5Trh}. When $P<P_{c}$, there are three branches on one $T-r_{h}$ line with fixed $P$. The two branches, i.e., the small and larger black hole branches, have positive slope, and thus have positive heat capacity $C_{P}$, which indicates these branches are thermodynamically stable. The slope of the other branch, i.e., the intermediate black hole branch, is negative, and this leads to negative $C_{P}$. This result is consistent with that shown in Fig. \ref{Pd5CP}. As analyzed in Section \ref{vanderWaalsfluid}, the three branches on one $T-r_{h}$ line with $P<P_{c}$ will construct a swallow tail behavior, and the first-order phase transition will occur. For clarity, the Gibbs free energy is shown in Fig. \ref{Pd5GT} for fixed $P$. It is clear that the swallow tail behavior appears when $P<P_{c}$. It is also worth noting that in Fig. \ref{PPrhgt}, the dashed lines correspond to negative $C_{P}$ implying a local thermodynamically unstable branch, while the solid ones correspond to positive $C_{P}$. The horizon radius $r_{h}$ increases from left to right along the $G-T$ line with fixed $P$.

\begin{figure}
\center{\includegraphics[width=8cm,height=6cm]{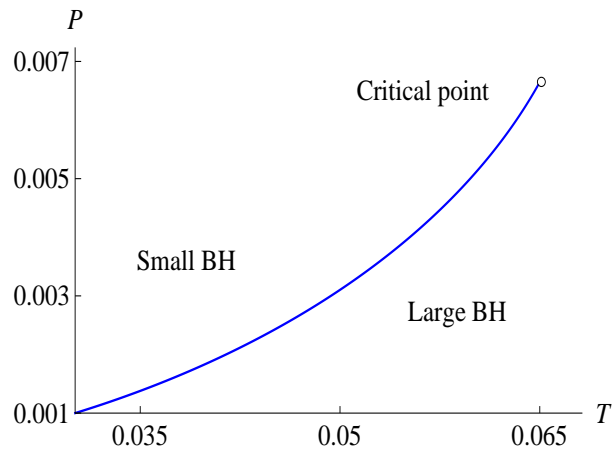}}
\caption{Phase diagram for the five-dimensional charged GB black hole with $Q=0.18$. The other value of $Q$ shares the same behavior of the phase diagram.}\label{Pd5PT}
\end{figure}

Now let us focus on the $G-T$ line with $P=0.0060<P_{c}$. If we start increasing the temperature from, say $T=0.062$, the system follows the small stable black hole branch (with smaller Gibbs free energy) until it approaches the intersection with the larger stable black hole branch, which corresponds to a SBH/LBH phase transition of first order at the intersection. Further increasing $T$, the system will follow along the larger stable black hole branch. So we observe the SBH/LBH phase transition. The situation is clearly illustrated in the $P-T$ diagram in Fig. \ref{Pd5PT}. There is a SBH/LBH line of coexistence, and it corresponds to the liquid/gas line of the van der Waals fluid. With the increasing of $T$, the coexistence line terminates at a critical point, at which the system undergoes a second-order phase transition.

We close this section by noting that there exists a SBH/LBH phase transition for a five-dimensional charged GB-AdS black hole, which is reminiscent of the liquid/gas phase transition of van der Waals fluid.

\section{The triple points and phase diagram when $d=6$}
\label{Phase2}

\begin{figure}
\center{\includegraphics[width=8cm,height=6cm]{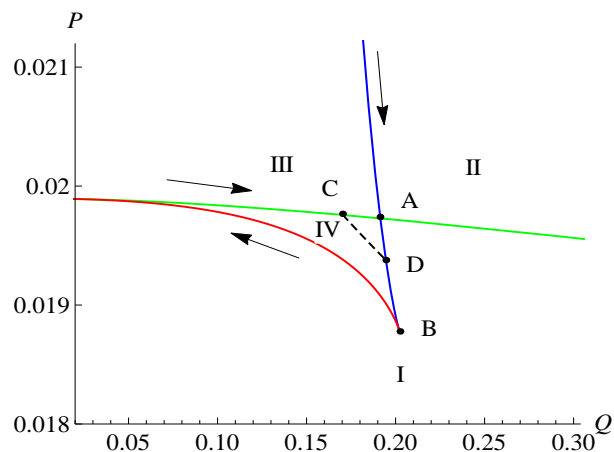}}
\caption{The critical point displayed in $P-Q$ plane in six-dimensional spacetime. The points $A$, $B$, $C$, and $D$ have coordinates (0.1914, 0.0197), (0.2018, 0.0188), (0.1705, 0.01976), and (0.1946, 0.0194). The blue, green, and red lines correspond to different critical points. These lines divide the $P-Q$ plane into four regions: I, II, III, and IV, in which the equation $(\partial_{r_{h}}T)_{P}=0$ has two roots, no root, two roots, and four roots. The black dashed line segment $CD$ denotes the triple point for different values of the charge $Q$, which is discussed in the text. The black arrows indicate increasing $r_{h}$.}
\label{Pd6PQ}
\end{figure}

\begin{figure}
\center{\subfigure[]{\label{Pd6q0gta}
\includegraphics[width=8cm,height=6cm]{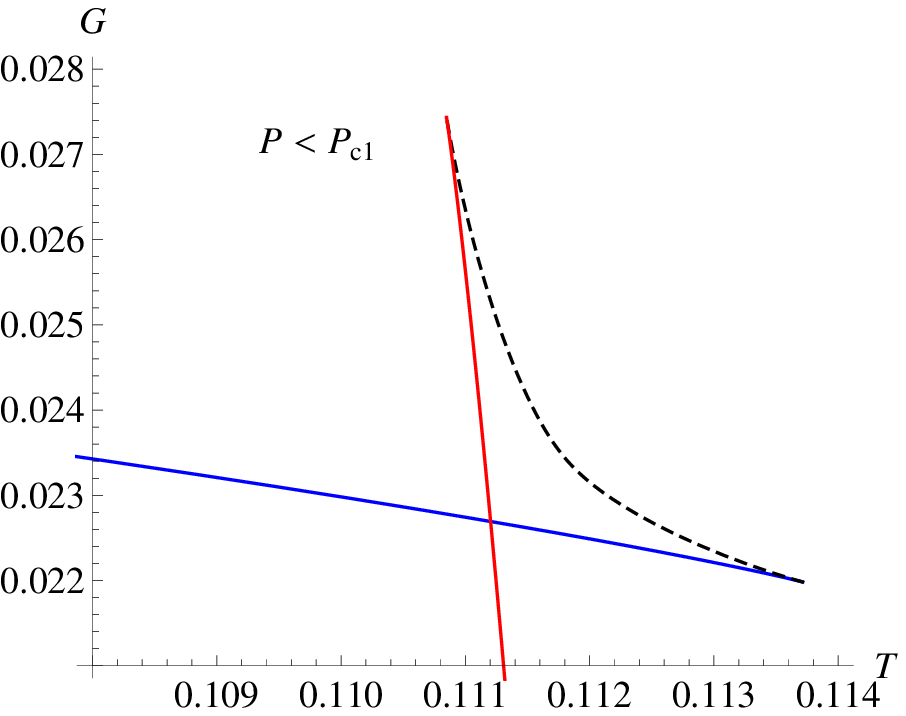}}
\subfigure[]{\label{Pd6q0gtb}
\includegraphics[width=8cm,height=6cm]{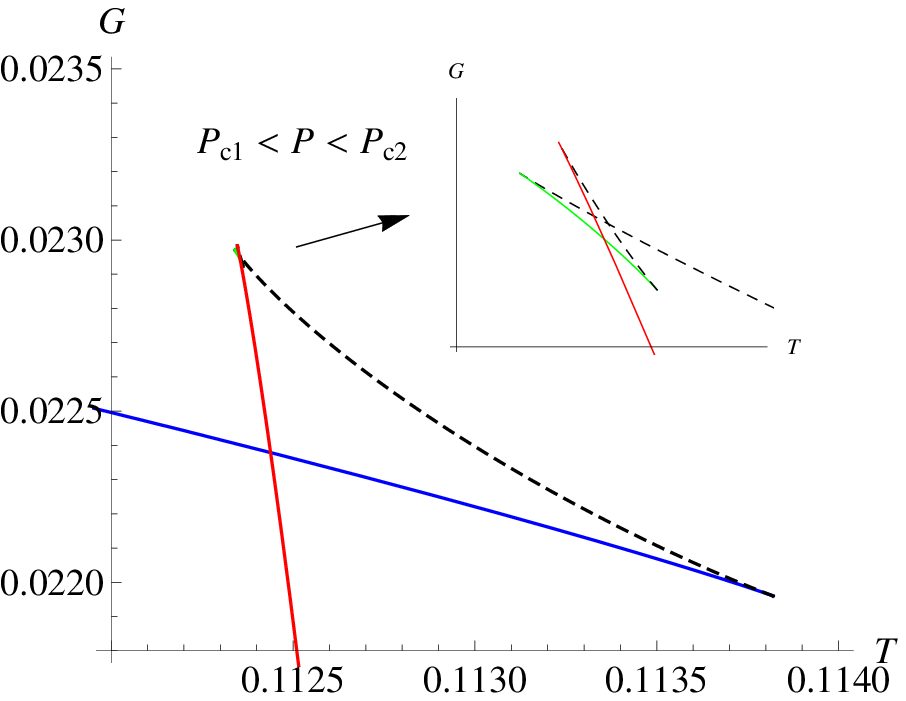}}}\\
\center{\subfigure[]{\label{Pd6q0gtc}
\includegraphics[width=8cm,height=6cm]{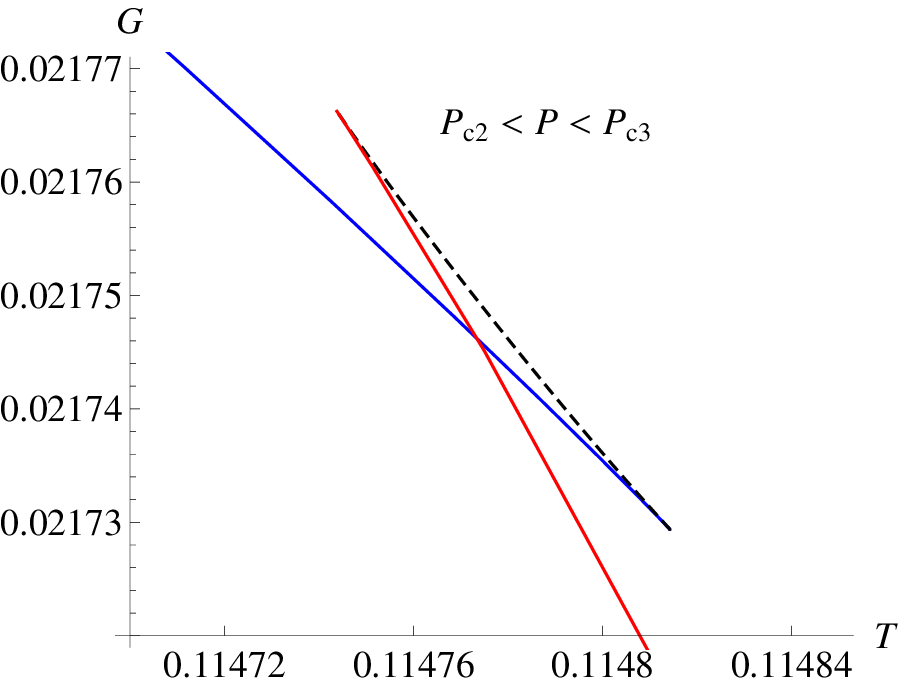}}
\subfigure[]{\label{Pd6q0gtd}
\includegraphics[width=6cm,height=5cm]{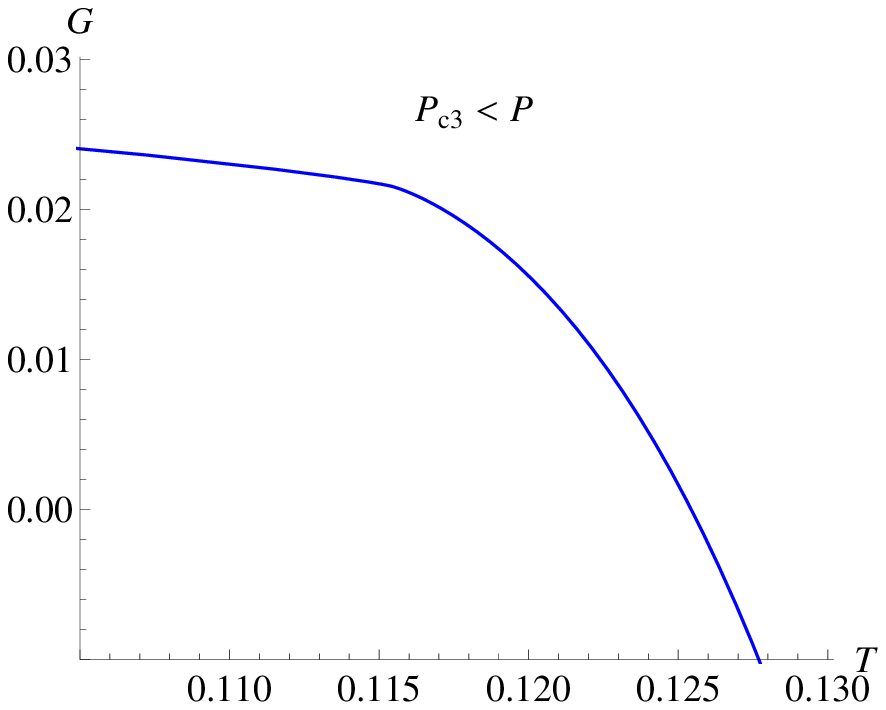}}}
\caption{Behavior of the Gibbs free energy $G$ in six-dimensional spacetime as a function of $T$ for the case of $Q=0.15<Q_{C}$.  The fixed pressure $P$ is taken as 0.0185, 0.0197, 0.03, and 0.035, respectively. The branches described by the black dashed lines are thermodynamically unstable. And these described by the blue, red, and green solid lines are stable ones. The critical points have the pressure $P_{c}=(0.01956, 0.01978, 0.03213)$. Other thermodynamic quantities at the critical points can be found in Table \ref{criticalparameters}. The horizon radius $r_{h}$ increases from left to right along the $G-T$ line.}\label{Pd6q0gtt}
\end{figure}

For a six-dimensional black hole, the temperature, Gibbs free energy, and heat capacity are
\begin{eqnarray}
 T&=&\frac{8\pi Pr_{h}^8+6r_{h}^6+2r_{h}^4-Q^2}{8\pi r_{h}^7+16\pi r_{h}^5},\\
 G&=&\frac{5Q^2\left(7r_{h}^2+20\right)-6r_{h}^4\left(4\pi Pr_{h}^6
     +(48\pi P-5) r_{h}^4+5r_{h}^2-20\right)}{480\pi r_{h}^3 \left(r_{h}^2+2\right)},\\
 C_{P}&=&\frac{r_{h}^2 \left(r_{h}^2+2\right)^2 \left(8\pi Pr_{h}^8-Q^2+6r_{h}^6
      +2r_{h}^4\right)}{2r_{h}^4\left(4\pi Pr_{h}^6+3(8\pi P-1)r_{h}^4
       +3r_{h}^2-2\right)+Q^2 \left(7 r_{h}^2+10\right)}.
\end{eqnarray}
For this case, the critical point can also be found with the condition (\ref{vdwcp3}). However, the situation is more subtle. In an appropriate range of the parameters, there are three critical points corresponding to SBH, IBH, and LBH, respectively. In the small charge limit, the universal relation for these critical points becomes,
\begin{eqnarray}
 \frac{P^{S}_{c}r^{S}_{hc}}{T^{S}_{c}}&=&\frac{\sqrt{5}}{16Q}-\frac{79}{48}
               +\frac{95839}{4608\sqrt{5}}Q-\frac{18057}{512}Q^{2}+\mathcal{O}(Q^{3}),\\
 \frac{P^{I}_{c}r^{I}_{hc}}{T^{I}_{c}}&=&\frac{1}{4}-\sqrt{\frac{11}{384}}Q
               -\frac{21}{64}Q^{2}+\mathcal{O}(Q^{3}),\\
  \frac{P^{L}_{c}r^{L}_{hc}}{T^{L}_{c}}&=&\frac{1}{4}+\sqrt{\frac{11}{384}}Q
               -\frac{21}{64}Q^{2}+\mathcal{O}(Q^{3}).
\end{eqnarray}
The behavior of the critical point is displayed in the $P-Q$ plane in Fig. \ref{Pd6PQ}. It behaves very different from that of the van der Waals fluid and the five-dimensional black hole case. There are four regions in the parameter space: I, II, III, and IV. In the regions I and III, $(\partial_{r_{h}}T)_{P}=0$ has two roots. In the region II, the equation has no root. And in the region IV, four roots will be obtained, which will result in a rich phase transition phenomena.

\begin{figure}
\center{\subfigure[]{\label{Pd6q0PQ}
\includegraphics[width=8cm,height=6cm]{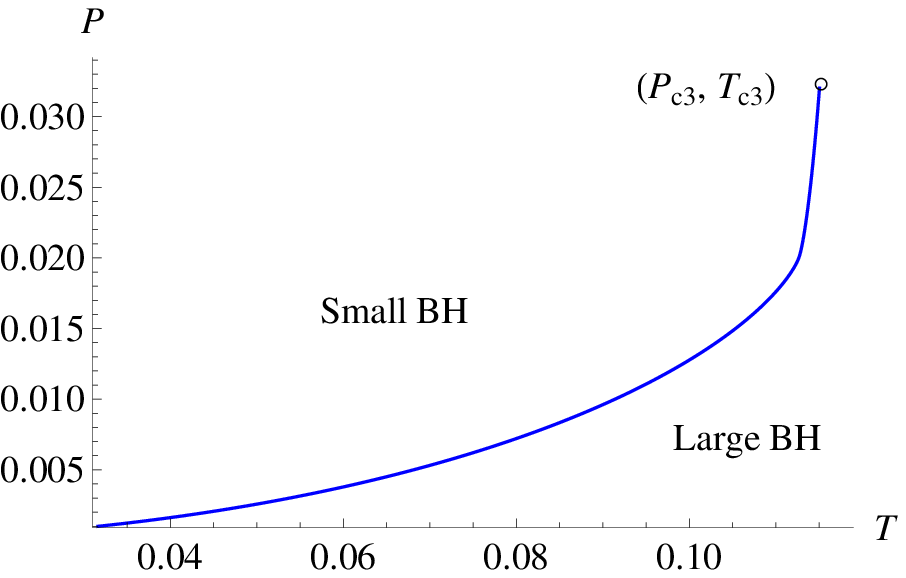}}
\subfigure[]{\label{Pd6Trh}
\includegraphics[width=8cm,height=6cm]{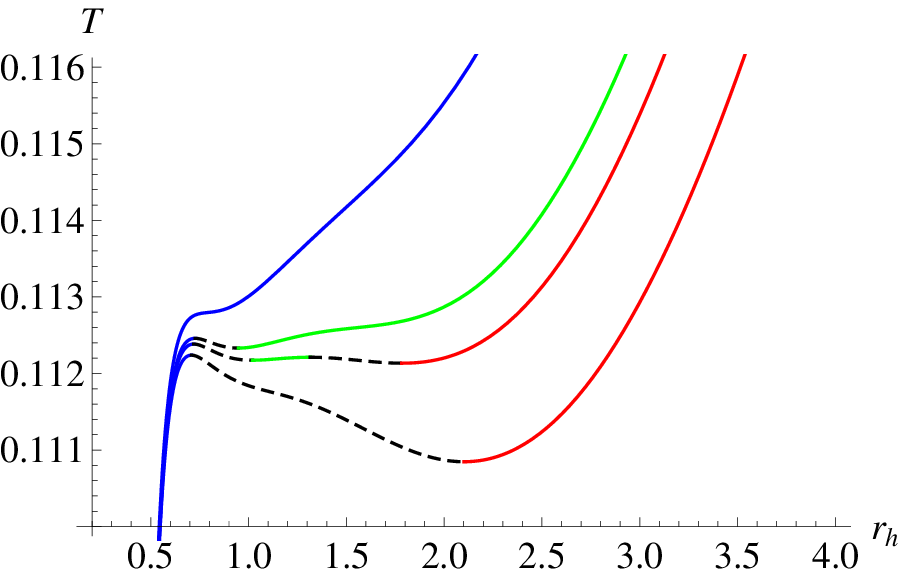}}}
\caption{Phase diagram and behavior of $T$ as a function of $r_{h}$ for the six-dimensional charged GB-AdS black holes. (a) Phase diagram for the case $Q=0.15<Q_{C}$. The critical point is located at $(P_{c3}, T_{c3})=(0.03213, 0.11868)$, which can also be found in Table \ref{criticalparameters}. (b) Behavior of $T$ as a function of $r_{h}$ for fixed $Q_{C}<Q=0.18<Q_{D}$ and $P$=0.0185, 0.0195, 0.02, 0.022 from bottom to top. The black dashed lines have negative slopes and describe unstable branches. For small pressure, we clearly see there are stable SBH and LBH branches, and unstable IBH branch. With the increasing of the pressure, there appears a new stable IBH branch. As we further increase the pressure, this branch disappears.}
\end{figure}

\begin{figure}
\centerline{\subfigure[$P=0.0185$]{\label{Pd6gta}
\includegraphics[width=8cm,height=6cm]{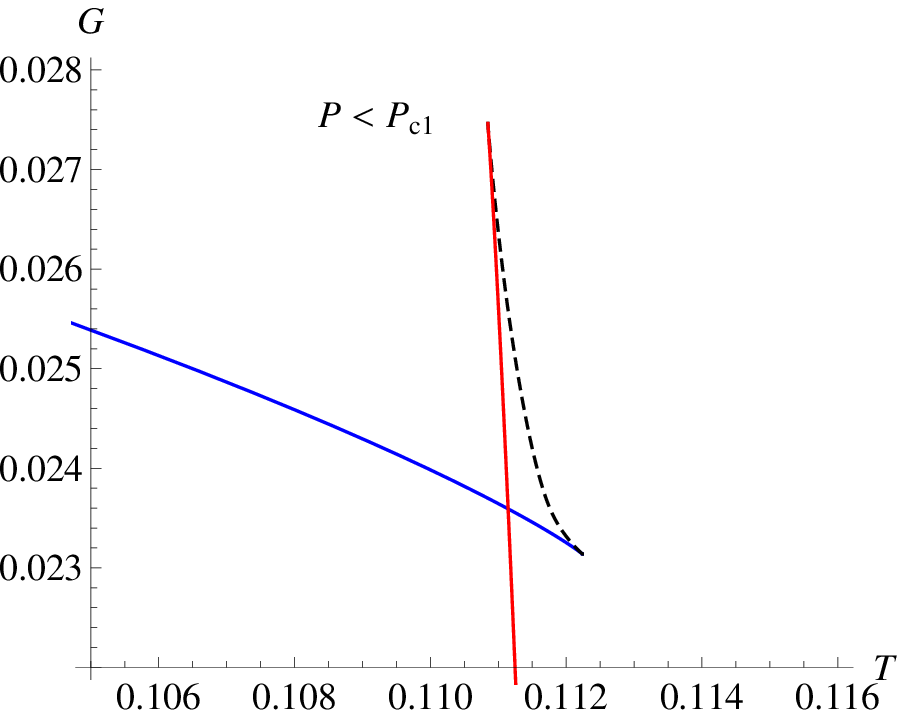}}
\subfigure[$P=0.0195$]{\label{Pd6gtb}
\includegraphics[width=8cm,height=6cm]{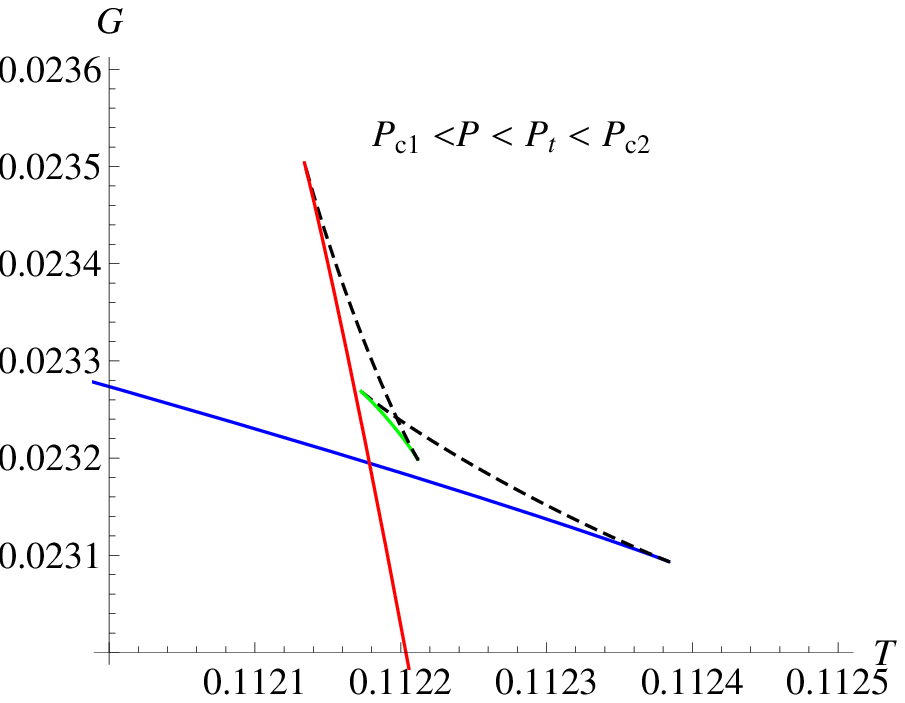}}}
\centerline{\subfigure[$P=0.019603$]{\label{Pd6gtc}
\includegraphics[width=8cm,height=6cm]{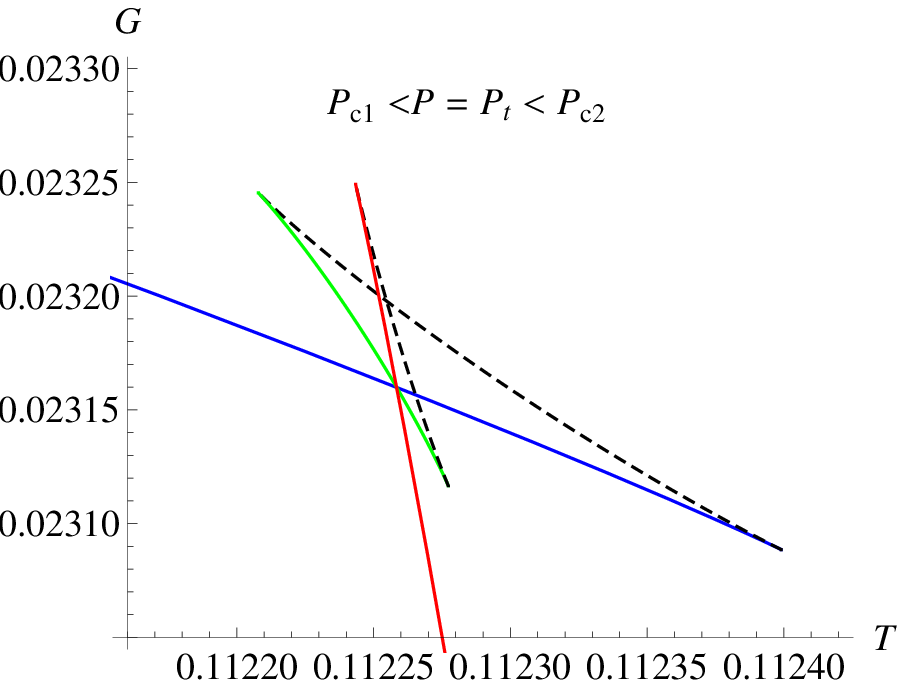}}
\subfigure[$P=0.01964$]{\label{Pd6gtd}
\includegraphics[width=8cm,height=6cm]{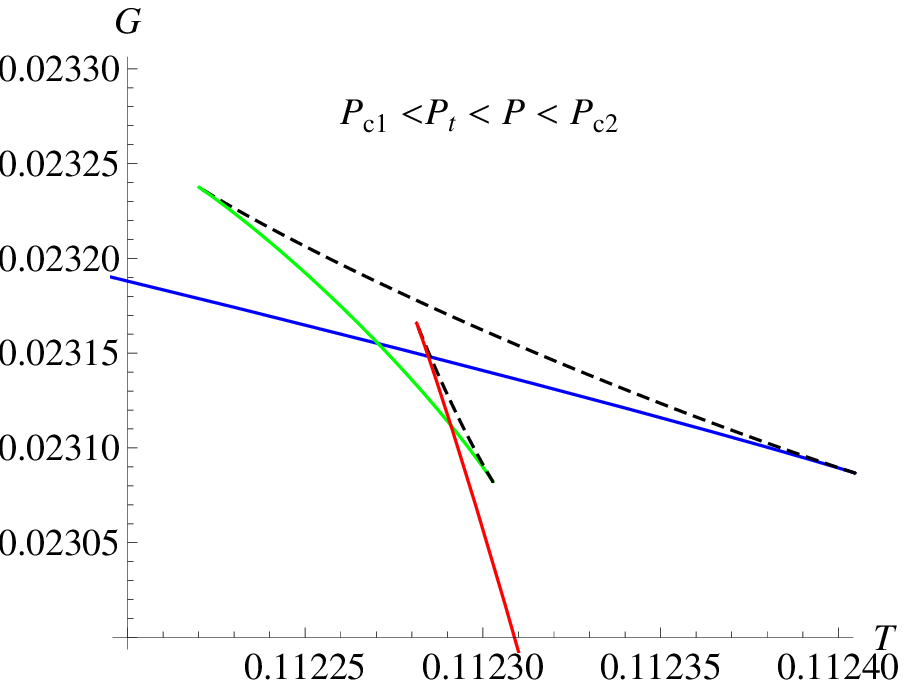}}}
\centerline{\subfigure[$P=0.02$]{\label{Pd6gte}
\includegraphics[width=8cm,height=6cm]{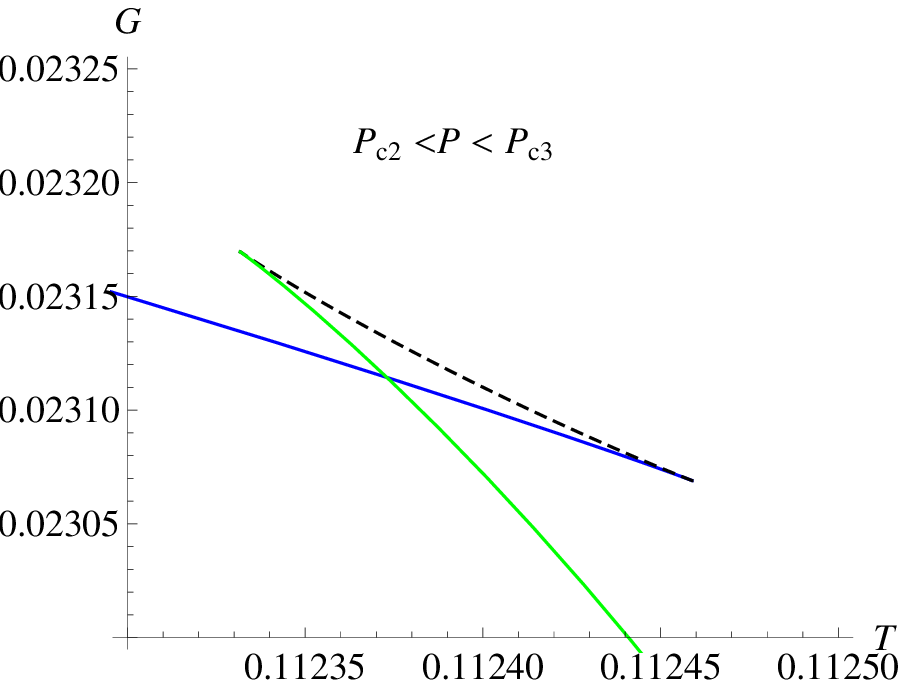}}
\subfigure[$P=0.022$]{\label{Pd6gtf}
\includegraphics[width=8cm,height=6cm]{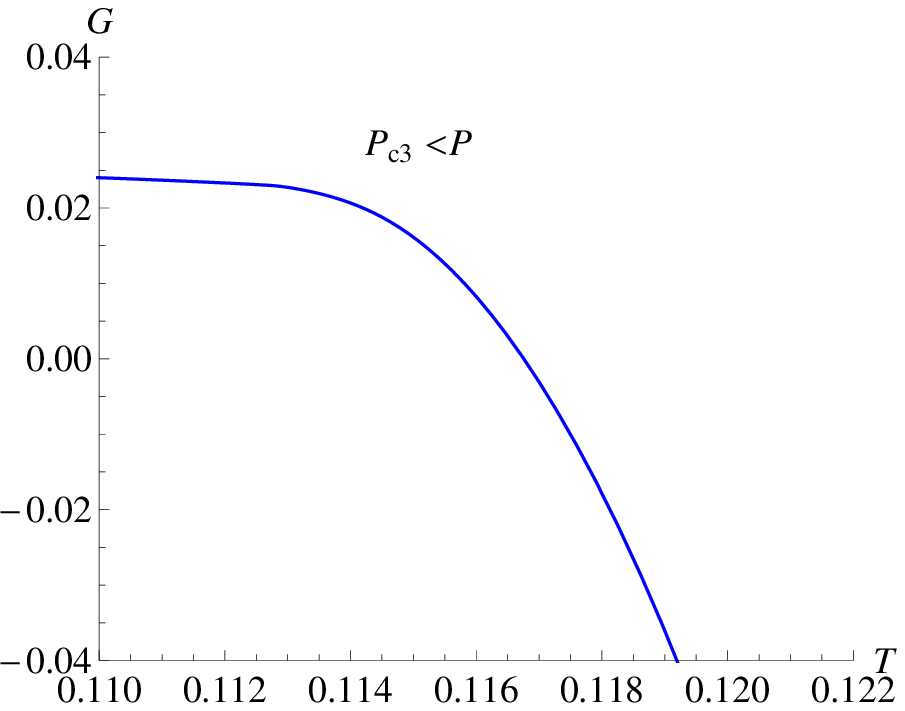}}}
\caption{Behavior of the Gibbs free energy $G$ as a function of $T$ for the six-dimensional charged GB-AdS black holes. The charge $Q$ is fixed as $Q_{C}<Q=0.18<Q_{D}$. The pressure $P$ is fixed as $P$=0.0185, 0.0195, 0.019603, 0.01964, 0.02, and 0.022. The unstable branches are described by black dashed lines. When $P<P_{t}$, there only exists the SBH/LBH phase transition. While when $P_{t}<P<P_{c2}$, there is the SBH/IBH/LBH phase transition. Thus $P=P_{t}$ is a triple point, at which the SBH, IBH, and LBH branches can coexist. The critical pressures are $(P_{c1},P_{c2},P_{c3})=(0.01927, 0.01974, 0.02155)$. And the pressure at the triple point is $P_{t}=0.01960$. The horizon radius $r_{h}$ increases from left to right along the $G-T$ line.}\label{Pd6gtt}
\end{figure}

According to the structure of the critical point displayed in Fig. \ref{Pd6PQ}, the following discussion will be restricted to fixed charges, $Q<Q_{C}$, $Q_{C}<Q<Q_{D}$, $Q_{D}<Q<Q_{B}$, and $Q_{B}<Q$.

\subsection{\textbf{Case 1}: \quad $Q=0.15<Q_{C}$}

First, we study the small charge $Q<Q_{C}$ case. Without loss of generality, we take $Q=0.15$. In this case, there are three critical points at $P_{c}$=(0.01956, 0.01978, 0.03213). The Gibbs free energy $G$ is plotted in Fig. \ref{Pd6q0gtt}. For $P<P_{c1}=0.01956$, there are three BH branches. The SBH and LBH branches (solid lines) are stable, while the IBH branch (dashed line) is unstable. There is one swallow tail, which implies that, with the increasing of $T$, the system will undergo a SBH/LBH phase transition with fixed $P$. When the pressure crosses $P_{c1}$, a new swallow tail behavior appears for the appearance of a new stable IBH branch described by the green solid line in Fig. \ref{Pd6q0gtb}. Since such branch has a larger value of $G$, it does not affect the phase transition. Thus there is still the SBH/LBH phase transition. Further increase the pressure such that $P_{c2}<P<P_{c3}$, the second swallow tail behavior disappears. And the SBH/LBH phase transition ends when the pressure $P$ approaches $P_{c3}$, at which the first-order phase transition turns to be a second-order one. The phase diagram is clearly shown in Fig. \ref{Pd6q0PQ}, and the result is consistent with the above analysis we given. We also clear that when the parameter locates in the region IV shown in Fig. \ref{Pd6PQ}, there are two swallow tails. Although the second one does not participate in the phase transition, it shows a rich structure of the $G-T$ line. And for the case 2, we will show that this two-swallow tail behavior will lead to a rich structure of phase transition, i.e., there exists a triple point and the SBH/IBH/LBH phase transition will occur in the proper range of the parameters.

\subsection{\textbf{Case 2}: \quad $Q_{C}<Q=0.18<Q_{D}$}

For the case of $Q=0.18$, which satisfies $Q_{C}<Q<Q_{D}$, the pressure $P$ has three critical points, i.e.,$(P_{c1},P_{c2},P_{c3})=(0.01927, 0.01974, 0.02155)$. We plot the temperature $T$ as a function of $r_{h}$ for $P=$0.0185, 0.0195, 0.02, and 0.022 (from bottom to top) in Fig. \ref{Pd6Trh}. The locally thermodynamically unstable branch is described by the dashed line. When $P<P_{c1}$, there are three branches, the stable SBH and LBH branches, and the unstable IBH branch. This situation is similar to the $d=5$ case, which implies that there exists a characteristic swallow tail behavior in the $G-T$ diagram, and a SBH/LBH phase transition will occur. Further increasing $P$ such that $P_{c1}<P<P_{c2}$, there appears a new stable IBH branch, which splits the unstable IBH branch into two, thus there are five branches along a $T-r_{h}$ line. Further increasing $P$, the stable IBH branch will disappear when $P_{2}<P<P_{3}$, and only one stable branch survives when $P>P_{c3}$. The corresponding Gibbs free energy is illustrated in Fig. \ref{Pd6gtt}. In the ranges $P<P_{c1}$ and $P_{c2}<P<P_{c3}$, it displays one characteristic swallow tail behavior. When $P>P_{3}$, there is no such behavior, and no phase transition occurs. However, in the range $P_{c1}<P<P_{c2}$ shown in Figs. \ref{Pd6gtc}-\ref{Pd6gtf}, the situation becomes more subtle. There appear three stable branches and two characteristic swallow tails, which imply a rich structure of the phase transition. For fixed $P$ satisfying $P_{c1}<P<P_{z}$, there is a SBH/LBH phase transition with the increasing of the temperature. As $P$ approaches $P_{z}$, we see that the three stable branches intersect in a unique point. At that point, three black hole phases (i.e., small, large and intermediate black holes) coexist together. Therefore, we observe a triple point characterized by $(P_{t}, T_{t})=(0.01960, 0.11226)$ at $Q=0.18$. Slight above this pressure, the system will emerge a standard SBH/IBH/LBH phase transition with the increase of $T$. And such phase transition disappears when $P_{c2}$ is approached. One thing worthwhile to note is that when the pressure $Q=Q_{A}$, the Gibbs free energy will encounter one swallow tail and two swallow tails with the increasing of $P$. And the two swallow tails simultaneously disappear at $P=P_{A}$.

\begin{figure}
\center{\includegraphics[width=8cm,height=6cm]{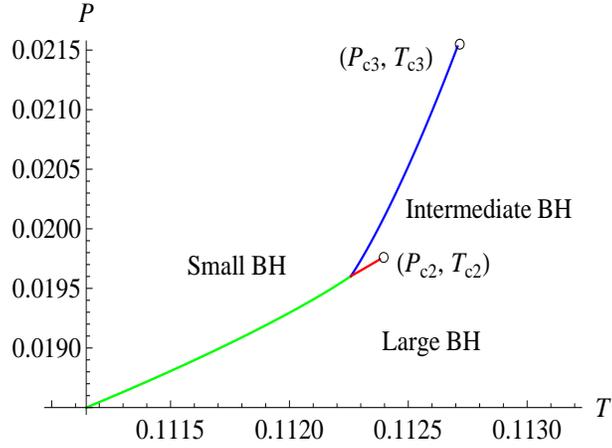}}
\caption{Phase diagram for the six-dimensional charged GB-AdS black hole with $Q_{C}<Q=0.18<Q_{D}$. Another value of $Q\in(Q_{C}, Q_{D})$ shares the same behavior of the phase diagram.}\label{Pd6PPTQ=0.18}
\end{figure}

The interesting phase diagram is displayed in Fig. \ref{Pd6PPTQ=0.18}. It is clear that there is a triple point, at which the small, large and intermediate black holes can coexist together. Below this pressure $P_{t}$, the system will undergo a SBH/LBH phase transition of first order with the increasing of $T$. Above $P_{t}$ and below $P_{c2}$, there will be a SBH/IBH/LBH phase first-order IBH/LBH phase transition ends at $P_{c2}$, at which the phase transition becomes the second order. And above the pressure $P_{t}$, the first-order SBH/IBH phase transition emerges and it terminates at critical $P_{c3}$. It is worthwhile to note that the critical point $(P_{c1}, T_{c1})$ is absent from this phase diagram. This is because that the first critical point measures the appearance of the second swallow tail, which does not participate in phase transition. Varying the charge $Q$, the triple point will be got, and it is shown in Fig. \ref{Pd6PQ} with the black dashed line $CD$. It is clear that the triple point is limited in a small range. The thermodynamic quantities at the triple point of different values of the charge $Q$ are listed in Table \ref{Triple}.

\begin{table}[h]
\begin{center}
\caption{The thermodynamic quantities at the triple point for different values of the charge $Q$.}\label{Triple}
\begin{tabular}{|c|c|c|c|c|c|c|c|c|}
  \hline
  \hline
   $Q$ &$P$ & $r_{h}$ & $v$ & $H$ & $T$&$S$&$\Phi$&$G$ \\
  \hline
   0.173 &0.01971&1.39476&1.05566&0.34787&0.11236&2.89145&0.00169&0.02299\\
   0.178 &0.01963&1.39484&1.05597&0.34784&0.11230&2.89190&0.00174&0.02308\\
   0.183 &0.01956&1.39492&1.05628&0.34782&0.11224&2.89234&0.00179&0.02317\\
   0.188 &0.01948&1.39500&1.05658&0.34779&0.11219&2.89278&0.00184&0.02326\\
   0.193 &0.01940&1.39508&1.05687&0.34777&0.11213&2.89321&0.00189&0.02336\\
  \hline\hline
\end{tabular}
\end{center}
\end{table}

There are some notes for the critical point and the triple point. The critical point implies that the black hole branch and the swallow tail behavior appear or disappear in the $T-r_{h}$ line or $G-T$ line. It is a multiple root of $(\partial_{r_{h}}T)_{P}=0$. Thus, the heat capacity $C_{P}$ diverges at that point. However, the triple point shown here is not consistent with the root of $(\partial_{r_{h}}T)_{P}=0$. So at the triple point, the heat capacity $C_{P}$ is finite and positive.

\subsection{\textbf{Case 3}: \quad $Q_{D}<Q=0.195<Q_{B}$}

\begin{figure}
\center{\subfigure[]{\label{Pd6q2gta}
\includegraphics[width=8cm,height=6cm]{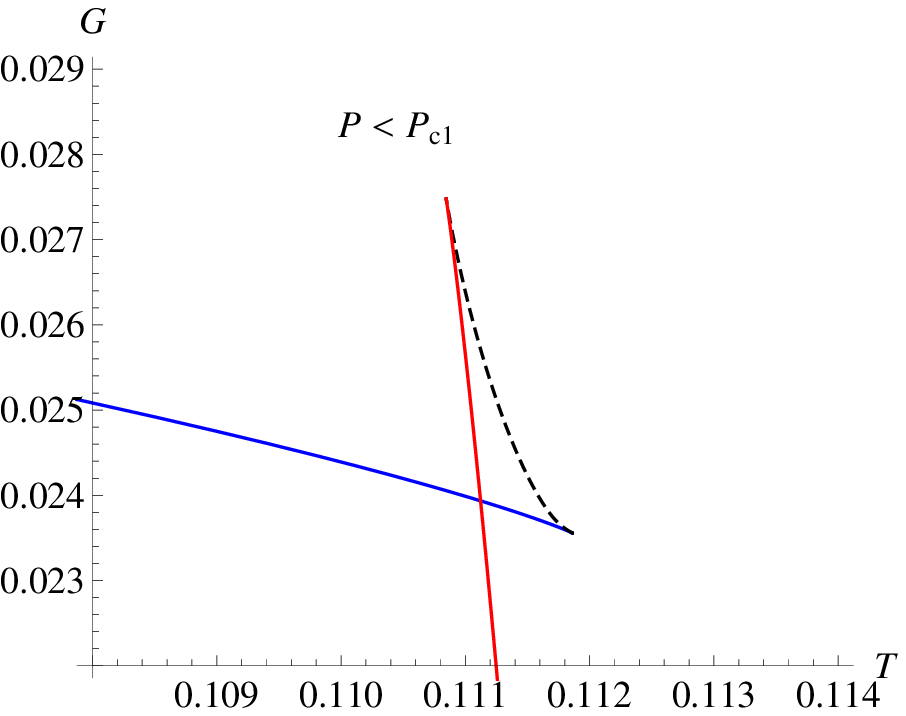}}
\subfigure[]{\label{Pd6q2gtb}
\includegraphics[width=8cm,height=6cm]{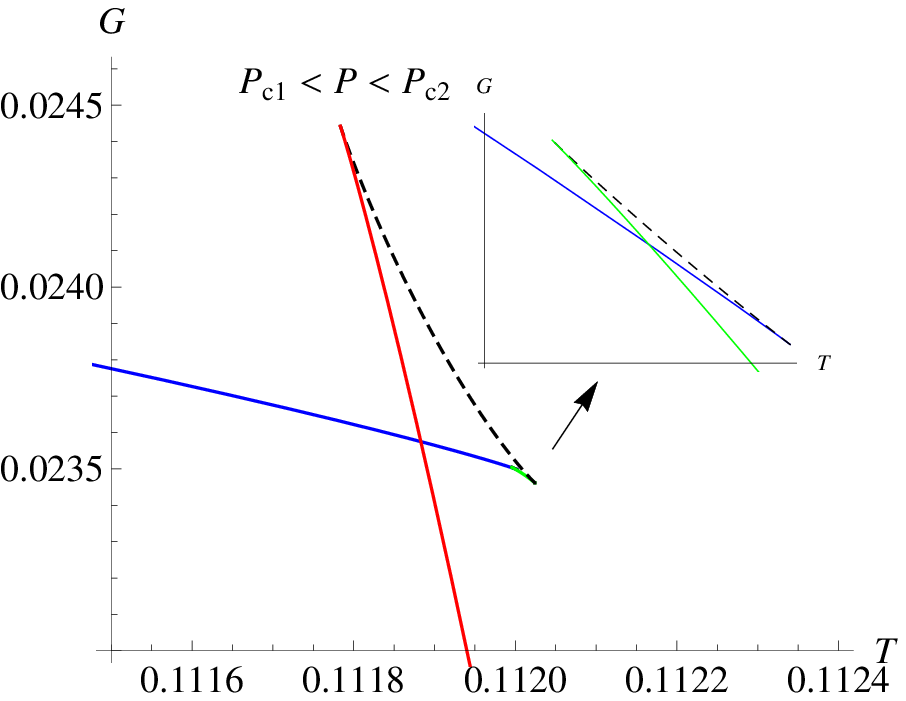}}}\\
\subfigure[]{\label{Pd6q2gtc}
\includegraphics[width=8cm,height=6cm]{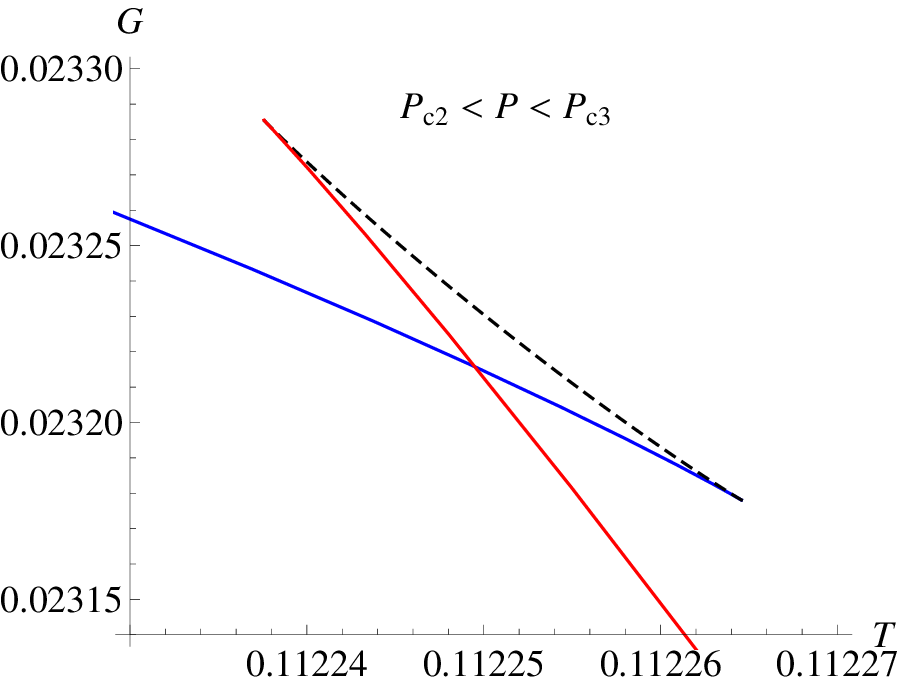}}
\subfigure[]{\label{Pd6q2gtd}
\includegraphics[width=8cm,height=6cm]{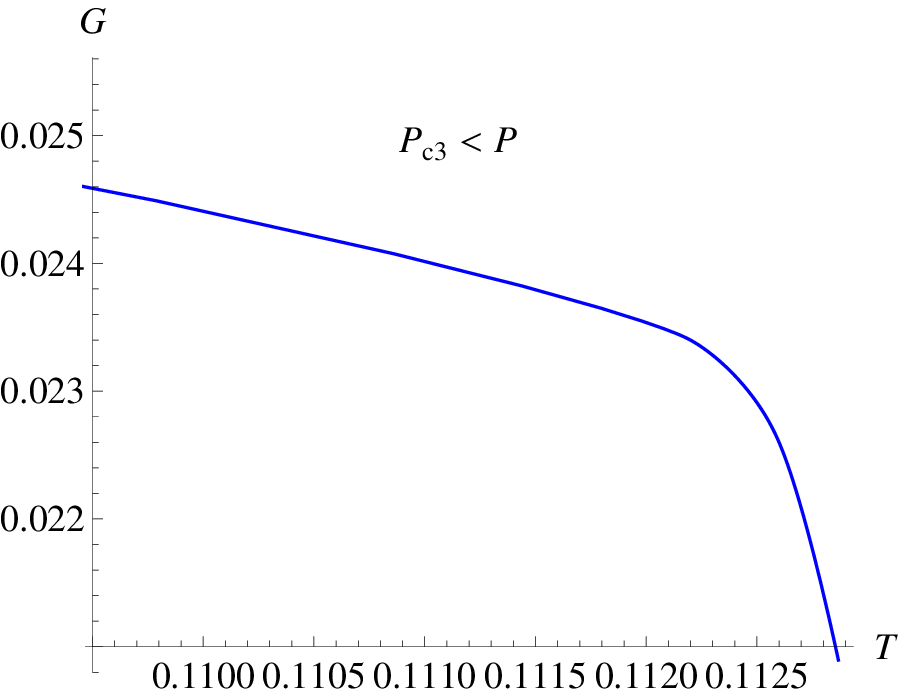}}
\caption{Behavior of the Gibbs free energy $G$ of the six-dimensional charged GB-AdS black hole as a function of $T$ for the case of $Q_{D}<Q=0.195<Q_{B}$. The pressure $P$ is fixed as $P$=0.0185, 0.0192, 0.0196, and 0.02. The unstable BH branches are described by the black dashed lines. For such case, only the SBH/LBH phase transition can be observed. The critical pressure is $(P_{c1},P_{c2},P_{c3})=(0.01900, 0.01933, 0.01972)$ . The horizon radius $r_{h}$ increases from left to right along the $G-T$ line.}\label{Pd6q2gt}
\end{figure}

For the case of $Q=0.195$ satisfying $Q_{D}<Q<Q_{B}$, the pressure $P$ also has three critical points, i.e., $(P_{c1},P_{c2},P_{c3})=(0.01900, 0.01933, 0.01972)$. The behavior of the temperature $T$ is similar to the case of $Q=0.18$. However, the behavior of the Gibbs free energy is different, which is depicted in Fig. \ref{Pd6q2gt}. When $P<P_{c1}$, there exists three black hole branches. The SBH and LBH branches are thermodynamically stable, while IBH branch is unstable. There exists a characteristic swallow tail for such case, indicating a SBH/LBH phase transition. When the pressure $P$ crosses the critical point $P_{c1}$, a new IBH branch appears, and there appears a second characteristic swallow tail. Further increasing the pressure such that $P_{c2}<P<P_{c3}$, the second swallow tail disappears. Finally, the first swallow tail fades out at $P=P_{c3}$. Since the stable IBH branch has a larger Gibbs free energy than the LBH branch at the same $P$ and $T$, it can be concluded that, in the range $0<P<P_{c3}$, we only observe the SBH/LBH phase transition with the increasing of $T$.

\begin{figure}
\center{\includegraphics[width=8cm,height=6cm]{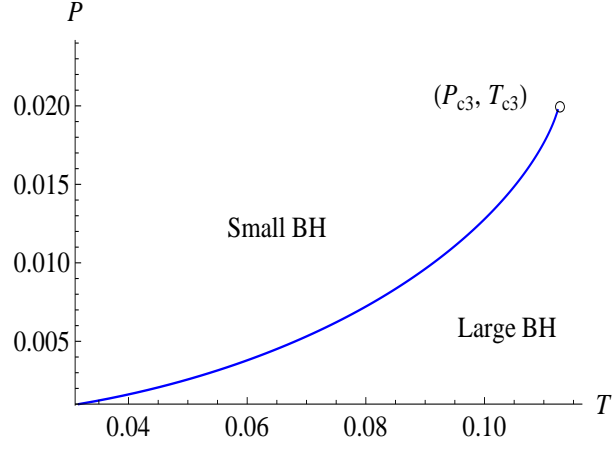}}
\caption{Phase diagram for the six-dimensional charged GB-AdS black hole with $Q_{D}<Q=0.195<Q_{B}$. Other values of $Q$ in the range $Q\in(Q_{D}, Q_{B})$ share the same behavior of the phase diagram.}\label{Pd6PPT}
\end{figure}

The phase diagram is illustrated in Fig. \ref{Pd6PPT}. At the coexistence line, the SBH and LBH phases have the same Gibbs free energy. Crossing this line with the increasing of $T$, the system will undergo a SBH/LBH phase transition of first order. The coexistence line terminates at a critical point $(T_{c3}, P_{c3})$, where the second-order phase transition occurs.

\subsection{\textbf{Case 4}: \quad $Q_{B}<Q=0.3$}

\begin{figure}
\center{\subfigure[]{\label{Pd6q3gta}
\includegraphics[width=8cm,height=6cm]{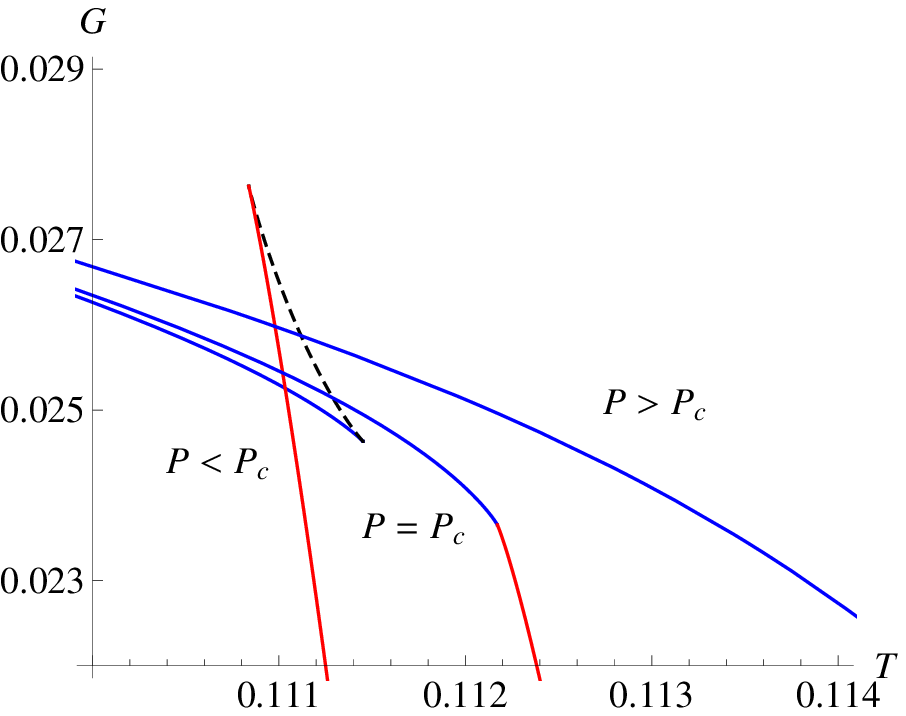}}
\subfigure[]{\label{Pd6q3PPT_12b}
\includegraphics[width=8cm,height=6cm]{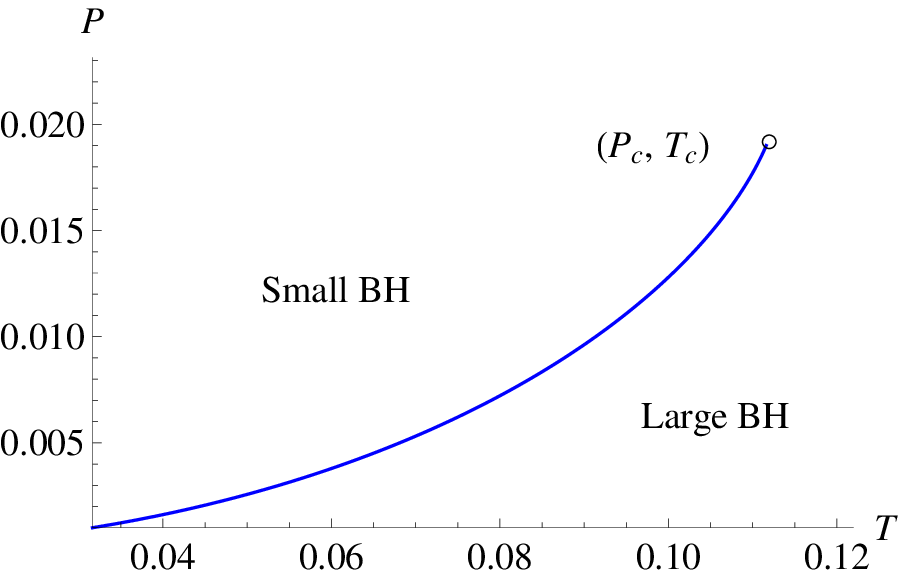}}}
\caption{Gibbs free energy and phase diagram for the six-dimensional charged GB-AdS black hole with $Q_{B}<Q=0.3$. (a) Behavior of the Gibbs free energy for fixed $P$=0.0185, 0.01956, and 0.025 form left to right. The horizon radius $r_{h}$ increases from left to right along the $G-T$ line. (b) Phase diagram. Other values of $Q$ in the range $Q_{B}<Q$ share the same behavior of the phase diagram.}\label{Pd6q3gt}
\end{figure}

There is only one critical point with $P_{c}=0.01956$ for the case of $Q=0.3(<Q_{B})$. It is quite similar to the van der Waals fluid. The SBH/LBH phase transition occurs for $P<P_{c}$ and it ends at $(T_{c}, P_{c})$, where the first-order phase transition becomes the second-order one, and the phase transition fades out for $P>P{c}$. The Gibbs free energy and the phase diagram are shown in Fig. \ref{Pd6q3gt}. This case is analogous to that of the van der Waals fluid.

\begin{figure}
\center{\subfigure[]{\label{Pd7pq}
\includegraphics[width=8cm,height=6cm]{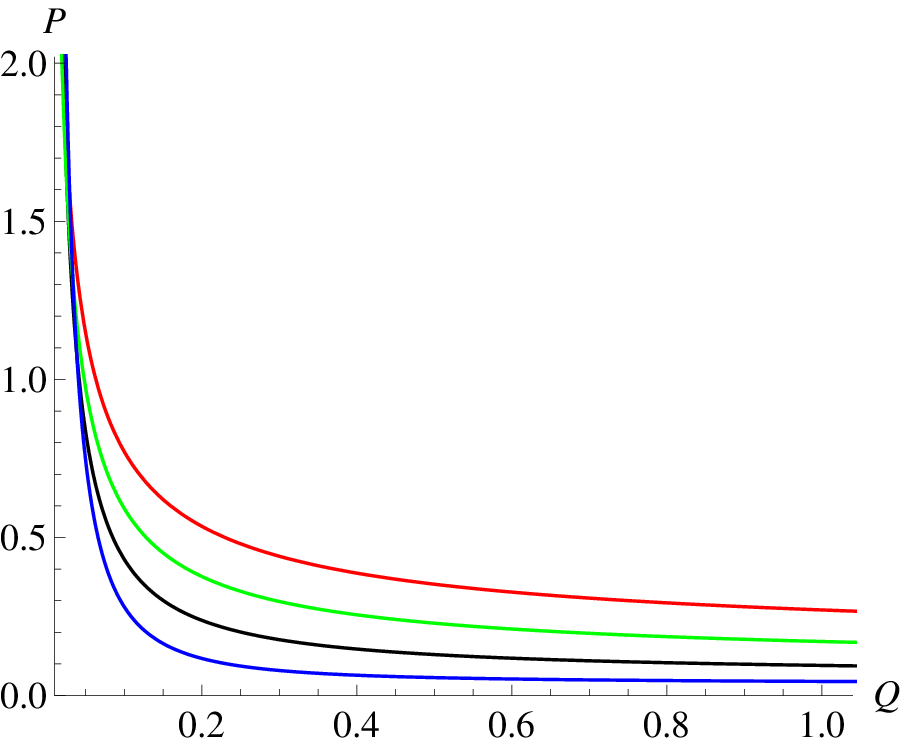}}
\subfigure[]{\label{Pd7pt}
\includegraphics[width=8cm,height=6cm]{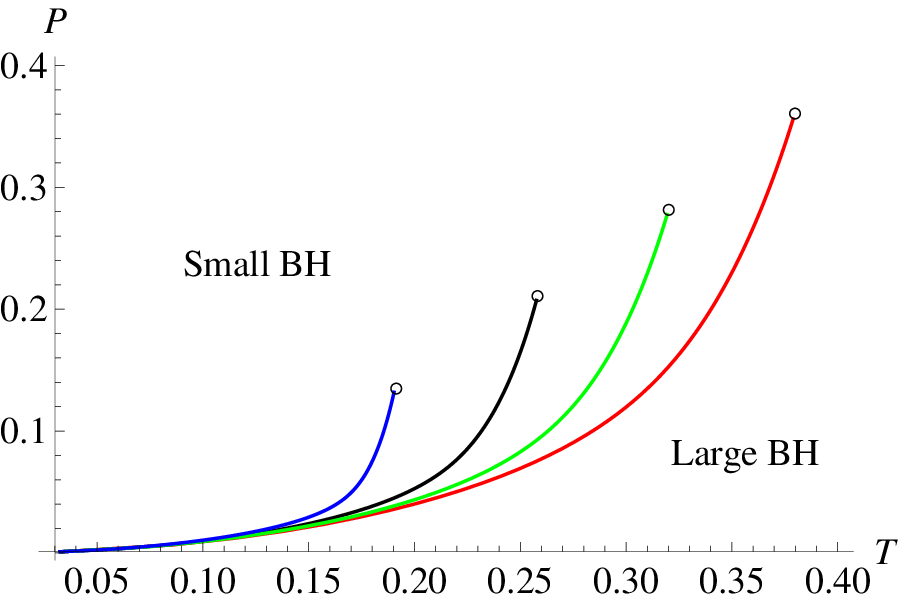}}}
\caption{(a) The critical point displayed in $P-Q$ plane for $d$=7, 8, 9, and 10 from bottom to top. (b) The phase diagram for the charged GB-AdS black hole with $d$=7, 8, 9, and 10 from left to right with fixed $Q=0.18$. It is clear that they have the similar behavior of the critical points and phase diagram.}\label{Pd77}
\end{figure}

\section{The phase diagram when $d\geq7$}
\label{Phase3}

When the dimension of the black hole is larger than 7, one naturally expects a richer structure of the phase diagram. However, it is not the case. We plot the critical point in the $P-Q$ plane in Fig. \ref{Pd7pq} for $d$=7-10. For fixed $d$, $(\partial_{r_h}T)_{P}=0$ has two roots below the line. While above the line, it has no root. Thus, it is clear that the behavior is quiet similar to the van der Waals fluid. And after some detailed study, it shows that there only the SBH/LBH phase transition. The triple point is not observed. The phase diagram is clearly illustrated in Fig. \ref{Pd7pt}, where only the SBH/LBH phase transition occurs. The values of these critical points are displayed in Table \ref{criticalparameters}.

The results show that the triple point only exists for $Q_{C}<Q<Q_{D}$ in $d=6$ spacetime. So it seems that $d=6$ is an exception. However, it is worthwhile to note that we have set the coupling constant $\alpha=1$. If we let it freely vary, one may find that the triple point could exist in $d\neq 6$ spacetime. However, in our opinion, the expectation of the triple point with $d\neq 6$ is still to some extent extremely weak.

\begin{table}[h]
\begin{center}
\caption{The values of these critical parameters shown in this paper. Note that all the critical points shown here have positive temperature, so these cases are corresponded to black hole system rather a non-black hole system.}\label{criticalparameters}
\begin{tabular}{|c|c|c|c|c|c|c|c|c|c|}
  \hline
  \hline
   $d$&$Q$ &$P$ & $r_{h}$ & $v$ & $H$ & $T$&$S$&$\Phi$&$G$ \\
  \hline
   5 &0.18 &0.00661&2.45821&9.12880&0.48082&0.06492&7.40091&0.00119&0.00037\\
      \hline
   6 &0.15 &0.01956&1.21313&0.52549&0.24906&0.11228&2.01315&0.00223&0.02303\\
     &     &0.01978&1.53448&1.70151&0.44343&0.11241&3.74069&0.00143&0.02295\\
     &     &0.03213&0.67641&0.52549&0.25578&0.11868&2.01315&0.00290&0.01687\\
   ~ &0.18 &0.01927&1.13426&0.37549&0.21392&0.11208&1.70034&0.00327&0.02333\\
     &     &0.01974&1.55453&1.81563&0.45861&0.11238&3.87652&0.00127&0.02297\\
     &     &0.02155&0.79278&0.06263&0.10495&0.11271&0.72726&0.00958&0.02298\\
   ~ &0.195&0.01900&1.06666&0.27617&0.18712&0.11193&1.46140&0.00426&0.02354\\
     &     &0.01933&0.87761&0.10412&0.12639&0.11203&0.91850&0.00765&0.02348\\
     &     &0.01972&1.56420&1.87280&0.46610&0.11236&3.94334&0.00135&0.02304\\
   ~ &0.3  &0.01956&1.62636&2.27567&0.51655&0.11217&4.39409&0.00185&0.02365\\
   \hline
   7 &0.18 &0.13225&0.73919&0.02719&0.08872&0.19038&0.39175&0.01199&0.01414\\
   \hline
   8 &0.18 &0.25874&0.77359&0.02368&0.09539&0.26462&0.32217&0.01034&0.01013\\ \hline
   9 &0.18 &0.40182&0.80394&0.02181&0.10533&0.33668&0.28934&0.00884&0.00791\\
    \hline
   10&0.18 &0.56449&0.82804&0.02033&0.11660&0.40754&0.27014&0.00767&0.00651\\
  \hline\hline
\end{tabular}
\end{center}
\end{table}

\section{Discussions and conclusions}
\label{Conclusion}

In this paper, we investigated the triple points and phase diagrams of the charged GB black holes in $d$-dimensional AdS space in the canonical (fixed charge $Q$) ensemble. We discussed this issue in the extended phase space with the cosmological constant treated as the pressure of the thermodynamic system and its conjugate quantity as the thermodynamic volume of the black holes.

We first dealt with the van der Waals liquid-gas system. By rewriting the equation of state into $T=T(v, P)$, we showed that the critical behavior and phase transition can also be equivalently determined by the $T-v$ line for fixed $P$. Then, by employing such treatment, we studied the phase diagram of the charged GB black holes in AdS space. It was shown that much information about the phase transition is encoded in the $T-r_{h}$ line, i.e., the divergence and sign of the heat capacity, critical point, number of the swallow tails of the Gibbs free energy, as well as the possible ranges where the SBH/LBH, SBH/IBH/LBH phase transitions take place. Thus, the $T-r_{h}$ line with fixed $P$ was shown to be a powerful tool to study the phase diagram of the black hole.

In this paper, we aimed to study the triple points and possible reentrant phase transitions in the charged GB-AdS black holes. And the method adopted here can be easily extended to other black holes. We summarize it as follows. First, plot the critical points determined by the condition (\ref{vdwcp3}) in the parameter space. Second, examine the number of the roots of $(\partial_{r_{h}}T)_{P}=0$ in each range of the parameter space. Using the number of the roots, we can get the possible type of the phase transition. For example, if there is only one root or no root, then where is no swallow tail behavior, and so no phase transition exists. And if there are two roots in some range of the parameter space, while no root in other range, then there exists the phase transition of the SBH/LBH type. If the number is larger than three, then there exists rich structure of the phase transition in some ranges of the parameter space, where the triple points and possible reentrant phase transitions may exist. Finally, combined with the behavior of the Gibbs free energy, the phase diagram will be obtained.

Using this method, we studied the phase diagram for the full range of the parameter charge $Q$. It was shown that the SBH/LBH phase transition exists in any dimension of $d$. However, in six-dimensional spacetime, the phase diagram is more subtle. In some range of the parameters, there are three critical points for a fixed charge $Q$, and $(\partial_{r_{h}}T)_{P}=0$ has four roots. In such range, the Gibbs free energy displays the behavior of two swallow tails. However, the triple point does not always exist in that range (i.e., in the range IV shown in Fig. \ref{Pd6PQ}). The triple point and the SBH/IBH/LBH phase transition are limited in the black dashed line $CD$ in Fig. \ref{Pd6PQ}. Thus we observe triple point and reentrant SBH/IBH/LBH phase transition in the charged GB-AdS black holes. However it seems that they are limited in a narrow range in parameter space.

Before closing the discussion, we need to stress another one issue. In this paper we only dealt with the case with fixed charge $Q$. One may wonder what happens when the pressure $P$ is fixed. In fact the situation is similar. This can be clearly interpreted from Fig. \ref{Pd6PQ} or other figures for the critical points displayed in $P-Q$ plane. The black hole system will encounter the similar structure of critical point for fixed $Q$ with the increasing of $P$ or for fixed $P$ with the increasing of $Q$. So the result is similar to the case with fixed charge $Q$.

We remark that in the charged GB-AdS black holes, we observed the existence of the triple point and the multiple first-order solid/liquid/gas phase transitions, and liquid/gas phase transitions of the van der Waals type, analogous to the ``every day thermodynamics".

\section*{Acknowledgements}
This work was supported by the National Natural Science Foundation of China (Grants No. 11205074 and No. 11375075), and the Fundamental Research Funds for the Central Universities (Grants No. lzujbky-2013-18 and No. lzujbky-2013-21).

\end{document}